\documentclass{aa}
\usepackage{amsmath}
\usepackage{txfonts}
\usepackage{amssymb}
\usepackage{esint}
\usepackage{natbib}
\usepackage{graphicx}
\usepackage{bm}
\usepackage[percent]{overpic}
\usepackage{multirow}
\usepackage[colorlinks=true, linkcolor=blue, citecolor=blue, filecolor=blue, urlcolor=blue]{hyperref}

\newcommand{\dif}{\mathrm{d}}

\begin{document}
\title{Shape index of Yarkovsky effect on irregularly shaped asteroids}
\author{Yining Zhang\inst{1,2}
	\and
	Yang-Bo Xu\inst{3}
	\and
	Zehua Qi\inst{1,2}
	\and
	Li-Yong Zhou\inst{1}\fnmsep\inst{2}
	\and
	Jian-Yang Li\inst{4}
}
\authorrunning{Zhang et al.}
\offprints{L.-Y. Zhou, \email zhouly@nju.edu.cn}
\institute{School of Astronomy and Space Science, Nanjing University, 163 Xianlin Avenue, Nanjing 210046, China
	\and
	Key Laboratory of Modern Astronomy and Astrophysics in Ministry of Education, Nanjing University, China
	\and 
	Shanghai Aerospace Control Technology Institute \& Shanghai Key Laboratory of Space Intelligent Control Technology, 1555 Zhongchun Road, Shanghai 201109, China
	\and 
	School of atmospheric science, Sun Yat-sen University, Zhuhai 519000, China
}

\date{}

\abstract{
The Yarkovsky effect on real asteroids is complicated to calculate either by analytical or numerical methods, since they are generally irregular in shape. We propose an index to properly characterise the shape of any asteroid, through which the Yarkovsky effect can be easily calculated without the heavy computations of surface temperatures. By analysing the energy absorbed and then emitted by a surface element, we find that the effective working power produced by the radiation recoil force on this surface element and its contribution to the Yarkovsky effect are both proportional to the double projected area of the surface element. The normalized total projected area over the asteroid's surface is defined as the shape index ($S_1$). We model the Yarkovsky effects of different asteroids using multiphysics software COMSOL, and take the rate of semi-major axis drift ($\dif a/\dif t$) obtained in these numerical simulations as the measurement of the strength of Yarkovsky effect. A linear relationship between $\dif a/\dif t$ and $S_1$ is confirmed. The shape index is then improved by taking the shadowing effect into account. A much better linear relationship is found between $\dif a/\dif t$ and the improved index $S_2$. This linear relationship is obeyed very well in a wide range of thermal parameter values. The influences of scattering and self-heating effects on the linear relationship are found ignorable. Using the shape index and the linear relation obtained in this paper, the rate of semi-major axis migration due to the Yarkovsky effect can be calculated accurately. Compared with the full numerical modeling of surface temperature and then the thermal radiation on an irregularly shaped asteroid, it is very easy to compute the shape index, which brings great convenience to the estimation of Yarkovsky effect. }

\maketitle

\section{Introduction}
\label{sec:intro}

A small rotating body orbiting around the Sun is heated by the solar radiation, and a specifically anisotropic temperature distribution appears on its surface. The absorbed energy is then released into space via thermal re-radiation, and the net recoil force of the thermal radiation may influence the asteroid's orbital evolution. 

Such a thermodynamic effect was first proposed by and later named after Polish Russian engineer Yarkovsky based on the erroneous ether theory in 1901, but it did not attract widespread attention \citep[see e.g.][for a brief review]{beekman2005nearly}. \citet{1951PRIA...54..165O} reintroduced this concept and noted that under the Yarkovsky effect, the semi-major axis of a small-sized object increases if it rotates progradely and decreases in the case of retrograde rotation. Then, after \citet{peterson1976source} proposed a two-step mechanism based on this effect to explain the origin of meteorites, the Yarkovsky effect came into focus. \citet{rubincam1987lageos} discovered the seasonal Yarkovsky effect while explaining the observed orbital decay of the satellite LAGEOS. Finally, the diurnal component (arising from the body's rotation) and the seasonal component (from orbital revolution) together constitute the complete Yarkovsky effect.

The Yarkovsky effect can be calculated either by analytical method or by numerical simulation. The former, often referred to as the ``linear theory'', is based on the assumption that the temperature at any location on the surface of a small object does not differ much from its average temperature. \citet{vokrouhlicky1998diurnal1} linearised the emission term ($\propto T^4$) in the equation of boundary condition, and obtained a complete linear solution for the diurnal Yarkovsky effect of a spherical body with an arbitrary spin axis obliquity. Later, this linear theory was extended to the biaxial ellipsoidal body \citep{vokrouhlicky1998diurnal2}. \citet{vokrouhlicky1999improved} reduced a small celestial body to a high-thermal-conductivity spheroid covered with a thin low-conductivity surface layer and obtained an analytical solution for the seasonal Yarkovsky effect. Finally, a complete solution that includes both the diurnal and seasonal components was proposed by \citet{vokrouhlicky1999complete}. 

Numerical computations of the Yarkovsky effect were developed with different thermal models for small bodies. Constrained by limitations in computing capability and observational accuracy, early thermal models for small celestial bodies often introduced certain simplified assumptions and approximations. The standard thermal model (STM) considers asteroids as smooth, non-rotating spheres, while the fast rotation model (FRM) is applicable to rapidly rotating asteroids with large thermal inertia \citep{1986Icar...68..239L,1989aste.conf..128L}. Both STM and FRM are primarily applicable to main-belt asteroids, while \citet{1998Icar..131..291H} established the near-Earth asteroid thermal model (NEATM). All these three models assume zero thermal emission on the night side of an asteroid. \citet{spitale2001numerical} and \cite{2002Icar..156..211S} used a finite-difference method to solve the three-dimensional (3D) heat conduction equation in an isotropic spherical body. They simultaneously calculated the diurnal and seasonal components, and their effects on the semi-major axis, eccentricity, and longitude of periapse. \cite{1996A&A...310.1011L} proposed the thermal physical model (TPM), approximating asteroids as triaxial ellipsoids and establishing a one-dimensional (1D) heat conduction model. Subsequently, \cite{1997A&A...325.1226L} further extended the model to irregularly shaped bodies, considering the effect of shadowing and self-heating. \citet{rozitis2011directional} proposed a well-developed advanced thermophysical model (ATPM), which incorporates the effects of shadowing, scattering of sunlight, self-heating and previously neglected thermal-infrared beaming caused by surface roughness. In ATPM, the lateral heat conduction is still neglected and only 1D heat conduction perpendicular to the surface is taken into account. Then \citet{rozitis2012rough} and \cite{rozitis2013global} adapted ATPM to both Gaussian random spheres and shapes of real asteroids to predict the semi-major axis drift caused by the Yarkovsky effect in the presence of surface roughness and global self-heating. \citet{basart2012modeling} modelled the 3D heat transfer incorporating the heat transfer and surface-to-ambient radiation modules in COMSOL\footnote{COMSOL Multiphysics$^\circledR$ \url{www.comsol.com}. COMSOL AB, Stockholm, Sweden.}. Using a sphere as an example, the study computed the dependence of the diurnal Yarkovsky effect on radius, rotation period and obliquity of incident radiation. \citet{xu2022diurnal} calculate the Yarkovsky effect of real asteroids and propose an index called ``effective area'' to measure the influence of asteroids' shape, which has been shown to be a fairly good index of the Yarkovsky effect on irregularly-shaped asteroids.

The linear theory gives an analytical solution for asteroids of regular shapes such as spheres and biaxial ellipsoids. Whereas for irregularly-shaped asteroids, we can either approximate them as regular shapes, which introduces certain errors, or numerically compute the surface temperature and then the recoil force of thermal irradiation, which gives more accurate results but a heavy calculation is required. The shape index can measure the effect of shape on the diurnal Yarkovsky effect and the application of such index allows a rapid and accurate estimation of Yarkovsky effect on arbitrarily-shaped asteroids. We notice that its applicability has not been clearly clarified, and its accuracy can be further improved. In this paper, we improve the definition of the shape index by taking into account the effects of projected shadows and self-heating, and we also investigate its applicability under different thermal parameters. The rest of this paper is organised as follows. In Section~\ref{sec:method}, we briefly review the heat conduction equation, and set up the numerical models. In Section~\ref{sec:shindx}, we define the shape index, improve it by taking into account the shadowing effect, and test the range of parameters for which the shape index works well. We also analyse the influence of lateral heat transfer, scattering and self-heating effects on the performance of the shape index. Finally we conclude our investigation in Section~\ref{sec:concls}.

\section{Model and methods}
\label{sec:method}
\subsection{Yarkovsky effect}
The Yarkovsky effect of an asteroid arises from the recoil force of the surface thermal radiation, which depends on the temperature distribution. The temperature in turn is determined by the heat conduction as well as the radiation absorbed and emitted by the surface. Since generally the diurnal Yarkovsky effect is much stronger than the seasonal effect \citep[see e.g.][]{vokrouhlicky1999complete, xu2020asteroid}, we focus on the former in this paper. In the following, the Yarkovsky effect simply refers to the diurnal effect, unless otherwise stated. 

In a rotating body-fixed coordinate system, the 3D heat conduction equation can be written as:
\begin{equation}\label{eq1}
\rho C \frac{\partial T}{\partial t}=K \mathrm{\nabla}^2 T,
\end{equation}
where $T$ represents the temperature, and $\rho$, $C$ and $K$ are the density, specific heat capacity, and thermal conductivity, respectively. The boundary conditions are then determined by 
\begin{equation}\label{eq2}
\epsilon\sigma T^4+K\left(\mathbf{n}\cdot\nabla T\right)=\alpha\mathcal{E},
\end{equation}
where $\epsilon$, $\sigma$, $\alpha$ and $\mathcal{E}$ are the emissivity, Stefan-Boltzmann constant, absorption coefficient, and external radiation flux, respectively, while $\mathbf{n}$ denotes the unit vector normal to the surface. Numerically, the heat conduction equation can be solved by the finite element method and the surface temperature is then obtained.
For a Lambert's isotropic thermal emission geometry, the recoil force generated by the small object's radiation is:
\begin{equation}\label{eq3}
\mathbf{F}=-\int{\frac{2}{3c}\epsilon\sigma T^4\mathbf{n}\dif S}.
\end{equation}

Substitute the force into the Gauss perturbation equation, we obtain the drift rate of the semi-major axis due to the Yarkovsky effect
\begin{equation}\label{eq4}
\frac{\dif a}{\dif t}= \frac{2a^{3/2}}{\sqrt{\mu\left(1-e^2\right)}} \left[ F_r e\sin\nu+F_t\left(1+e\cos\nu\right) \right],
\end{equation}
where $\mu=G\left(M_{\odot}+m\right)$ is the reduced mass of the Sun and the small body, $e$ is the orbital eccentricity, and $\nu$ the true anomaly. In addition, $F_r$ and $F_t$ are the radial and tangential components of the Yarkovsky force ($\mathbf{F}$) respectively.

Equation~\eqref{eq4} gives the instantaneous acceleration of the semi-major axis ($\dif a/\dif t$), which might change with time as the asteroid rotates and revolves along its orbit. Generally, the rotation period $P$ is much shorter than the revolution period $P_\text{orb}$, so it is reasonable to assume that the temperature distribution on the asteroid's surface reaches a state of dynamic equilibrium at any phase of its orbit. Therefore, the surface temperature can be calculated numerically and the corresponding Yarkovsky force can be averaged over one rotation period ($P$) at each position along the orbit. And the semi-major axis drift rate in long term should be averaged again over the revolution period ($P_\text{orb}$).

\subsection{Finite element method}
The linear theory of Yarkovsky effect works quite well for objects of regular shape, but it might bring considerable error if we approximately treat an irregular asteroid as an ellipsoid that the linear theory can handle \citep{xu2022diurnal}. In this paper, we numerically calculate the temperature of an asteroid of any shape using the software COMSOL, in which the heat transfer equation is computed in 3D model, that is, the heat conduction in both lateral and perpendicular directions to the asteroid's surface is taken into account. The model also incorporates shadowing, scattering of sunlight and self-heating effect. But the roughness-induced thermal-infrared beaming effect is not included because the surface is always set as Lambertian surface. 

Our main interest in this paper is to quantify the influence of an asteroid's shape on the Yarkovsky effect. Currently, the number of asteroids whose shapes are well-known is still limited. We obtain from the Planetary Data System\footnote{\url{https://sbn.psi.edu/pds/shape-models/}} 34 real asteroids as our samples of irregularly-shaped objects. For the sake of direct comparison, these shape models are then scaled to have the same volume as a sphere of radius $R=10$\,m. 

The triangular facet shape models of these samples are then constructed and fed the software. Requiring a reasonable accuracy of temperature, the resolution of these shape models is automatically set by COMSOL. Thousands of meshes for the surface of each model were generated. An asteroid is assumed to be a core composed of free tetrahedra covered by a shell of certain thickness. The heat exchange occurs barely not in the core, but only in the shell \citep[see e.g. ][]{xu2022diurnal}. Composed of triangular prisms, the shell is set to be 5 times thick as the thermal penetration depth $l_\text{d}$, and is then divided into 25 layers to make the mesh for numerical calculations. We note that the penetration depth, $l_\text{d}=\sqrt{KP/(2\pi\rho C)}$, is the e-folding depth at which the temperature variation amplitude is decreased by a factor of $1/e$. 

We set the asteroid at the initial moment as an isothermal object, with the temperature $T_\text{ini}$ determined by the balance between the outward thermal radiation and the absorption of solar radiation, that is, $4\pi R^2\cdot\epsilon\sigma T_\text{ini}^4=\pi R^2\cdot\alpha\mathcal{E}$. The time step used to calculate the variation of temperature is taken to be one hundredth of the rotation period ($P/100$) in the simulation. Generally, the surface temperature reaches a dynamic equilibrium after several cycles of rotation, then the temperature at any point on the surface varies periodically. When the difference of temperature distribution at a certain moment from the one at the same phase in the next rotational cycle is less than 0.1\%, the dynamic equilibrium state is regarded as reached. The number of rotation periods required for an asteroid to reach the equilibrium depends on thermal parameters. Particularly, an asteroid attains the equilibrium quickly when the value of thermal conductivity $K$ is small, and vice versa.

The actual thermal processes could be very complex. The thermal parameters are not constants but functions of temperature. The albedo may be a function of wavelength, which means that the surface reflects the self-heating radiation in infrared with a different efficiency from the Solar radiation that is described by the Bond albedo.  However, to simplify our analyses, we set in this paper the thermal parameters as constants and take the albedo in the infrared band as the Bond albedo. The symbols and values of the parameters we use in this article are summarised in Table~\ref{tab:para}.

\begin{table}
\centering
\caption{The symbols and values of the parameters in the numerical model.}
\label{tab:para}
\begin{tabular}{c c c}
\hline
\hline
Symbol & Description & Value \\
\hline
$a$ & semi-major axis & $1$\,AU \\
$\gamma$ & spin obliquity & $0^\circ$ \\
$P$ & rotation period & $1800$\,s \\
$\alpha$ & absorptivity & $0.9$ \\
$\epsilon$ & emissivity & $0.9$ \\
$\rho$ & densitiy of the regolith layer & 1500 \,kg\,m$^{-3}$ \\
$C$ & specific heat capacity & 680 \,J\,kg$^{-1}$\,K$^{-1}$ \\
$K$ & thermal conductivity &  \\ 
\hline
\end{tabular}
\end{table}

The rotation period $P=1800$\,s in Table~\ref{tab:para} is relatively short compared with the observed periods in common asteroids. We note that it does not affect our analyses in this paper, but it brings great convenience in our computation using COMSOL. When the thermal conductivity is low, at a fixed time step ($P/100$), a shorter rotation period serves to reduce errors in the calculation of temperature distribution. 
Additionally, for asteroids of large size ($R\gg l_d$), the Yarkovsky effect is determined by the ``thermal parameter'' $\Theta$ \citep{vokrouhlicky1998diurnal1} defined as
\begin{equation}
	\Theta = \frac{\Gamma\sqrt{\omega}}{\epsilon \sigma T_{\star}^3},
\end{equation}
where $\Gamma = \sqrt{\rho C K}$ is the thermal inertia, $\omega=2\pi/P$ is the angular velocity of the body's rotation, and $T_{\star}$ is the subsolar temperature defined by $\epsilon \sigma T_{\star}^4=\alpha \mathcal{E}$.
Therefore, the value of period $P$ in fact can be traded with other parameters like $\rho$ and $C$, and the impact of a too-short rotation period can be mitigated by adjusting other parameters. In this paper, we varied the thermal conductivity $K$ across four orders of magnitude, ensuring that the choice of rotation period does not compromise the reliability of our conclusions.

\section{Shape index}
\label{sec:shindx}

We have found in our previous paper \citep{xu2022diurnal} an approximately linear dependence of the diurnal Yarkovsky effect on the ``effective area'' of an irregularly shaped asteroid, in which the Yarkovsky effect was measured by the drift rate of semi-major axis $\dif a/\dif t$. And the effective area was used to asses the characteristics of an asteroid's shape that affects the Yarkovsky effect. Below we will briefly revisit the concept of the effective area and improve it by taking into account some geometrical and physical effects that were ignored previously. 

\subsection{Definition of shape index}
Since the rate of an asteroid's migration caused by the Yarkovsky effect depends explicitly on its spin obliquity $\gamma$ as $\dif a/\dif t \propto \cos \gamma$ \citep{vokrouhlicky1998diurnal1,xu2022diurnal}, we describe the strength of the Yarkovsky effect in the simplest case with the spin axis being perpendicular to the orbital plane, that is, $\gamma=0^\circ$.

For a polyhedral shape model consisting of $N$ facets, if the $i$-th surface element with an area $a_i$ is illuminated by the Sun at moment $t$, then the Solar radiation energy absorbed by the surface element at this moment is:
\begin{equation}\label{eq5}
E_{\rm in}=-\alpha\mathcal{E}a_i\left(\mathbf{n}_i \cdot\mathbf{n}_{\rm in}\right),
\end{equation}
where $\mathbf{n}_i$ and $\mathbf{n}_{\rm in}$ denote the directions of this surface element and the Solar incidence. We note that these directions are defined in a body-fixed frame centred on the asteroid, in which $\mathbf{n}_i$ is fixed and $\mathbf{n}_{\rm in}$ rotates. 

When the temperature distribution on the surface of an asteroid has reached the dynamic equilibrium, the temperature of any surface element should satisfy $T\left(t+P\right)=T(t)$, and the amounts of heat absorbed and released by a certain surface element should be equal in every rotation period. For an asteroid of size much larger than the penetration depth ($R\gg l_{\rm d}$), the heat budget of any surface element is approximately completed locally but not globally, that is, only in the same surface element. 
Ignoring the heat exchange between neighbouring surface elements, a surface element absorbs the Solar radiation, stores the energy in it, and then re-radiates the energy from the same surface element after a delay of time $\delta t$. Given the rotation frequency $\omega$, the delay in time $\delta t$ means that the re-radiation will be omitted after the surface element rotates by an delay angle $\theta=\omega \delta t$. Therefore the recoil force produced by the thermal radiating of energy absorbed by this surface element at this moment is in the direction $\mathbf{n}_{\rm rf} = -\mathbf{n}_i e^{{\rm i}\theta}$, where ${\rm i}=\sqrt{-1}$ and $e^{{\rm i}\theta}$ denotes a rotation by angle $\theta$. Since the delay angle $\theta$ is determined by the thermal parameter $\Theta$, for a homogeneous asteroid, we expect that all surface elements have the same $\delta t$ and therefore the same delay angle $\theta$. 

In our model, the Solar incidence always lies on the equatorial plane of the body, thus the tangential direction of the circular orbit is $\mathbf{n}_{\rm t} = \mathbf{n}_{\rm in} e^{{\rm i}\pi/2}$. Finally, the magnitude of the recoil force projected on the tangential direction produced by this surface element at this moment is proportional to the energy released in the same direction:
\begin{equation}\label{eq6}
E_{\rm out}\propto -\left(\mathbf{n}_{\rm rf}\cdot\mathbf{n}_{\rm t}\right)E_{\rm in} \propto \alpha\mathcal{E}a_i\left(\mathbf{n}_i\cdot\mathbf{n}_{\rm in}\right) \left( \mathbf{n}_i e^{{\rm i}\theta} \cdot \mathbf{n}_{\rm in} e^{{\rm i}\pi/2}\right).
\end{equation}
The net recoil force of the thermal radiation produced by the $i$-th surface element in a rotation period can be obtained by averaging over a rotation. 

If the shape of an asteroid is fully convex, there will be no projected shadows and self-heating effect. Each surface element is exposed to the Solar radiation for exactly half rotation, and the average contribution of each facet to the Yarkovsky effect over a rotation period is proportional to:
\begin{equation}\label{eq:average_t}
F_{t,i}\propto\frac{1}{P}\int_{t_0}^{t_0+P/2} \alpha\mathcal{E}a_i \left(\textbf{n}_i\cdot\textbf{n}_{\rm in}\right)\left( \mathbf{n}_i e^{{\rm i}\theta} \cdot \mathbf{n}_{\rm in} e^{{\rm i}\pi/2} \right)\dif t, 
\end{equation}
where $t_0$ is the sunrise moment of the $i$-th surface element. Denote all unit vectors by direction cosines.  Particularly, the Solar incident is in $\mathbf{n}_{\rm in}=(\cos\beta_x, \cos\beta_y, \cos\beta_z)$. In the body-fixed frame, the direction cosine can be reduced to $\mathbf{n}_{\rm in}=(\cos\beta, \sin\beta, 0)$, where $\beta$ is the angle between the $x$-axis and $\mathbf{n}_{\rm in}$. The average over time in a rotation period in Eq.\,\eqref{eq:average_t} is equivalent to an average over the angle $\beta$, and further algebraic calculations show
\begin{equation}\label{eq:average_ang}
	\begin{aligned}
F_{t,i}\propto & \frac{1}{\pi}\int_{-\pi/2}^{+\pi/2} \alpha\mathcal{E}a_i \left(\textbf{n}_i\cdot\textbf{n}_{\rm in}\right)\left( \mathbf{n}_i e^{{\rm i}\theta} \cdot \mathbf{n}_{\rm in} e^{{\rm i}\pi/2} \right)\dif \beta \\
 & = \frac{1}{\pi}\alpha\mathcal{E}a_i \sin\theta \int_{-\pi/2}^{+\pi/2}  \left(\textbf{n}_i\cdot\textbf{n}_{\rm in}\right)^2 \dif \beta \\
 & = \frac{1}{4} \alpha\mathcal{E}a_i \sin\theta \left(\textbf{n}_i\cdot\textbf{n}_{\rm noon}\right)^2.  
\end{aligned}
\end{equation}
Obviously, in this integration, we have set $\beta=0$ when this surface element is at its local noon, and $\mathbf{n}_{\rm noon}=\mathbf{n}_{\rm in}(\beta=0)$. 

The contributions from all surface elements can be summarised to represent the overall Yarkovsky force. The summation can be normalized by dividing the corresponding value obtained from a spherical asteroid of the same thermal parameters and with the same bulk volume. The constant ``parameters'' including $\alpha$, $\mathcal{E}a_i$ and the delay angle $\sin\theta$, are removed in the normalization, and finally a shape index $S_1$ is defined as such a normalized summation. In fact, it's the ratio of the ability of an object with an arbitrary shape to produce the Yarkovsky effect to that of a sphere with the same volume and thermal parameters, and it reads: 
\begin{equation}\label{eq:s1}
	S_1 = \frac{A}{4}\sum^N _i a_i \left(\textbf{n}_i\cdot\textbf{n}_{\rm noon}\right)^2,
\end{equation}
where $A=\left[6/(\pi V^2)\right]^{1/3}$ and $V$ is the volume of the asteroid.  Surely, according to the definition, $S_1=1$ for a spherical asteroid.  The index $S_1$ describes qualitatively the influence of shape on the Yarkovsky effect. We note that $S_1$ has the same physical essence as the ``effective area'' defined in \citet{xu2022diurnal}, only except the former has been normalised and thus is dimensionless, while the latter has been proved to serve well as an index of the relative strength of Yarkovsky effect of irregularly shaped asteroids. 

\begin{figure*}[!htbp]
	\centering
	\begin{overpic}[width = 9cm]{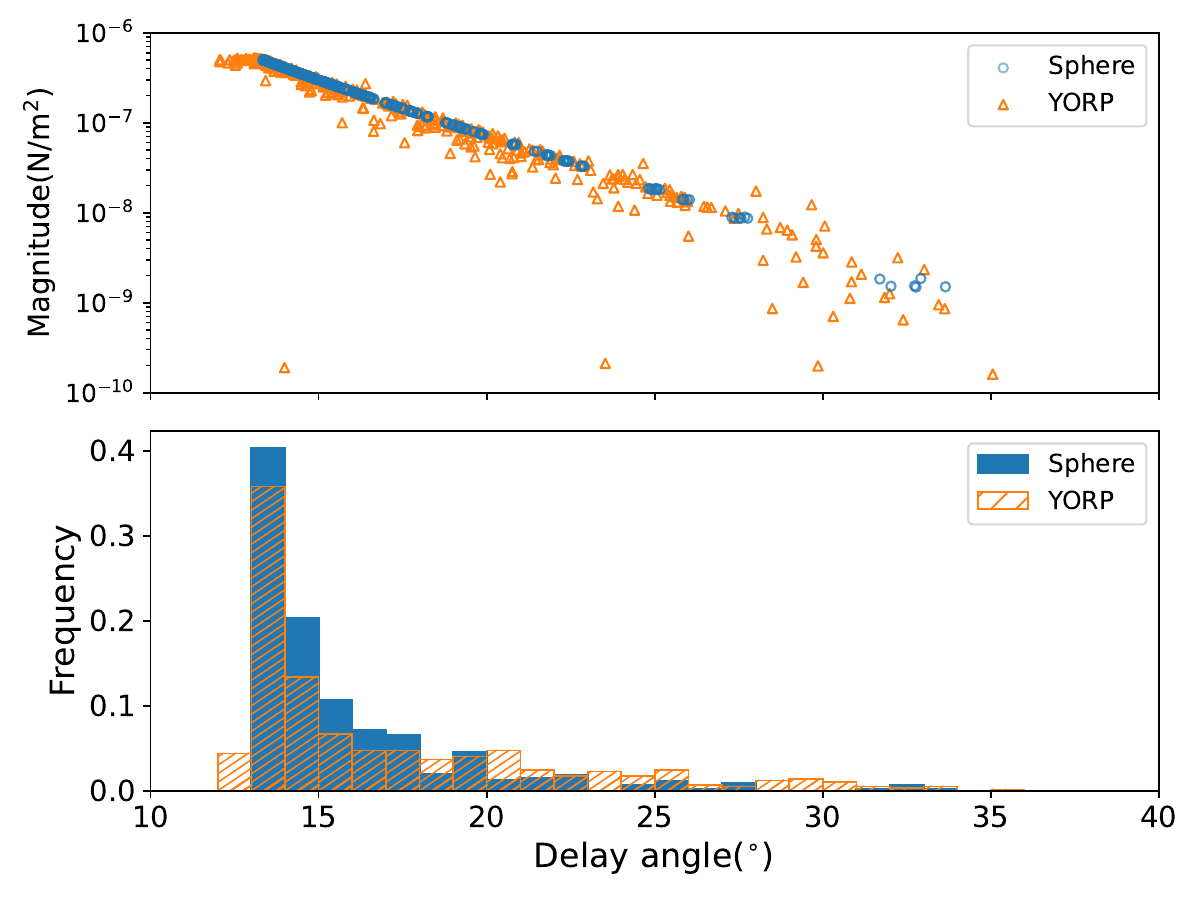}
		\put(1,73){(a)}
		\label{a}
	\end{overpic}
	\hspace{0mm}
	\begin{overpic}[width = 9cm]{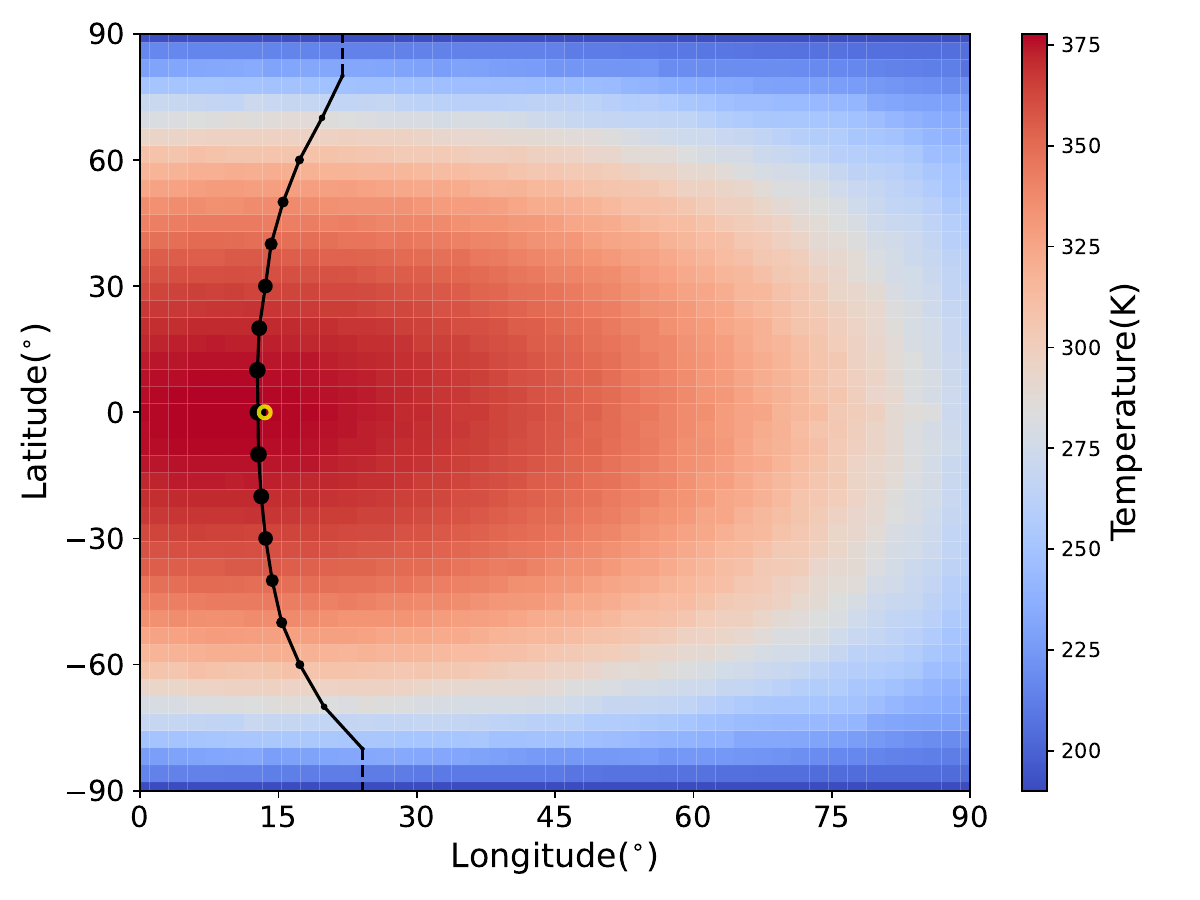}
		\put(1,73){(b)}
	\end{overpic}
	
	\caption{The directions and magnitudes of  the averaged radiation forces of surface elements on sphere model and YORP model. A thermal conductivity $K_1=0.0015$\,Wm$^{-1}$K$^{-1}$ (correspondingly $\Theta = 0.736$) is adopted for both models. The direction and magnitude have been projected in the orbital (equatorial) plane here (see text). (a) The magnitudes of radiation forces of surface elements versus the delay angle (upper), and the frequency of delay angles of all surface elements (lower). The delay angle of a surface element is the angle between the radiation force averaged over a rotation period and the corresponding averaged force of the same surface element but with $K = 0$. Blue and orange indicate the results for sphere and YORP model, respectively.  (b) The direction and magnitude of the radiation force at different latitudes on the surface of a sphere. The longitude $0^{\circ}$ is at noon. The longitude of the black dot at certain latitude indicates that the radiation force at this latitude points to the direction given by the longitude. The size of black dots represents the magnitude of the radiation force (in N/m$^2$, see text) projected in the orbital plane. The yellow circle indicates the direction and magnitude ($F/4\pi R^2$) of the total radiation force. The background colour gives the surface temperature. }
	\label{fig:angle}
\end{figure*}

So far, we have assumed that the delay ($\delta t$ in time or $\theta$ in angle) between the absorption and emission of energy on all surface elements are the same on an asteroid. To verify this assumption, using COMSOL we numerically calculate the surface temperature of two model asteroids, namely, a spherical asteroid and the asteroid YORP (scaled to have the same volume as a sphere of radius 10\,m). The delay occurs all the time, but in practice, it is not easy to define it at each moment. Instead, we calculate the radiation force of each surface element and average it over a rotation period. As comparison, we set the heat conductivity $K=0$ (therefore no heat conduction and no delay at all) and calculate the same averaged force from the same surface element. Then the delay angle is defined as the angular difference between directions of the averaged forces in these two cases. We plot the statistics of delay angles of two asteroid models in Fig.~\ref{fig:angle}. 

Since the force component perpendicular to the orbital (equatorial) plane is not of interest in this paper, the delay angles and the magnitudes of the force shown in Fig.~\ref{fig:angle} are in fact their projections on the equatorial plane. We note that considering different sizes of facets used in calculations, we use the pressure (force per area) instead of the force to represent the magnitude of radiation force in Fig.~\ref{fig:angle}. With the same thermal parameters for both sphere and YORP model, regardless of their shapes, the delay angles of surface elements from both models display a similar distribution. Most of the delay angles are concentrated at $13^\circ\sim14^{\circ}$ as shown in Fig.~\ref{fig:angle}(a) where thermal parameter $\Theta = 0.736$ is adopted. In addition, these facets with nearly the same delay angle have the largest radiation pressures and contribute the most to the Yarkovsky effect. Although the delay angles of some surface elements located in the polar region or greatly affected by the shadowing effect (in the YORP model) deviate significantly from  $13^\circ\sim14^{\circ}$, their radiation pressures are very low, and they contribute very little to the Yarkovsky effect. 

More intuitively, we show in Fig.~\ref{fig:angle}(b) the surface temperature distribution on the sphere model and the radiation forces at different latitudes. Because of the geometric symmetry of the sphere model, all facets at a certain latitude on a sphere are identical. Thus the direction of the radiation force at different longitudes tells directly the delay angle (from the subsolar point).  From Fig.~\ref{fig:angle}(b), we see that the delay angles in the polar region are larger than those in low latitudes. However, the magnitudes of the radiation forces at high latitude region are much smaller than the ones at low latitude region. Therefore, the total radiation force are mainly determined by the forces produced in the equatorial region, and their delay angles are identical to each other. So, the assumption that all surface elements of a body have the same delay angle $\theta$ determined by the thermal parameter $\Theta$, is valid.  

In the definition of shape index $S_1$ in Eq.\,\eqref{eq:s1}, the illumination of a surface element is determined only by its own direction, which is true only if the body is fully convex. But,small bodies of metres to kilometres in radii, which are most affected by the Yarkovsky effect, are likely to have irregular shapes due to their weak self-gravity, and they might not be simply approximated as fully convex. Particularly, many of them tend to have non-negligible projected shadows, under which the surface area apparently has lower temperature and thus weaker thermal radiation. Taking into account the shadowing effect, we may improve the shape index as follows. 

For surface elements that are in the projected shadows of other elements at a given moment $t$, they are unable to absorb energy from the Solar radiation, and thus they are assumed to have barely  thermal radiation and contribute negligibly to the Yarkovsky force before the Sun rises (locally for them). Therefore, a correction to Eq.\,\eqref{eq5} can be made as:
\begin{equation}
\label{eq9}
E_{\rm in}\propto -\alpha\mathcal{E}a_i\big[1-s_i(t)\big] \left(\textbf{n}_i\cdot\textbf{n}_{\rm in}\right),
\end{equation}
where $s_i(t)$ indicates whether the $i$-th surface element is in shadow at time $t$, with $s\left(t\right)=1$ if it is and $s\left(t\right)=0$ if it is not.
Following the same derivation as shape index $S_1$, a corrected shape index $S_2$ can be defined as:
\begin{equation}\label{eq:s2}
S_2 = \frac{A}{P}\int_{0}^{P} \Bigg[\sum^N_i a_i\big(1-s_i(t)\big) \big(\textbf{n}_i\cdot\textbf{n}_{\rm in}\big)^2\Bigg]\dif t.
\end{equation}
To calculate $S_2$, we need to run the shadow tests and compute the integration over a rotation period.

As we have shown above, the power of thermal radiation in the direction of orbital velocity is proportional to the shape index. We note that the drift rate of semi-major axis ($\dif a/\dif t$)  of an asteroid is actually proportional to the changing rate of the mechanical energy (the power) of an orbit, which in turn is driven by the recoil force of the thermal radiation. Therefore, a linear relationship between the Yarkovsky effect (measured by $\dif a/\dif t$) and the shape index is expected.

\subsection{Linear relation}
\label{sec:linear}

The relationship between the index $S_1$ and the strength of the diurnal Yarkovksy effect has been investigated by \citet{xu2022diurnal}, in which a linear dependence of the semi-major axis drift rate on $S_1$ has been verified using 34 shape models of real asteroids.

When the shape of an asteroid is known, the indices $S_1$ and $S_2$ can be easily computed according to their definitions in Eq.\,\eqref{eq:s1} and Eq.\,\eqref{eq:s2}. While the surface temperature of an asteroid with given thermal parameters can be calculated using COMSOL and thus the Yarkovsky effect (represented by the drift rate of semi-major axis) can be obtained. All 34 models adopted in \citet{xu2022diurnal} have been assumed to spin around the $z$-axis. It is true that most of asteroids do rotate around the principal axis (that is the $z$-axis) except for those tumblers.  However, to extend the sample size of shape models, in this paper we assume that these asteroids also rotate around the $x$-axis and $y$-axis (while the Sun is always on the equatorial plane). Since rotating around different axes produces different Yarkovsky effects as well as different shape indices, by this way we finally have 102 ($=34\times 3$) shape models. 

We adopt the model of asteroids as described in Section \ref{sec:method} with parameters given in Table~\ref{tab:para}. A thermal conductivity $K=0.0015$\,Wm$^{-1}$K$^{-1}$, which is the typical value for the regolith-covered asteroids \citep{Farinella1998}, is adopted here. With these parameters, the thermal parameter $\Theta= 0.736$ and the scaled radius of the body $R'=R/l_{\rm d}\sim 10^4$, and an asteroid experiences approximately 50 rotation periods before the thermal equilibrium is reached across the body. 

\begin{figure*}[!htbp]
	\centering
	\begin{overpic}[width = 9cm]{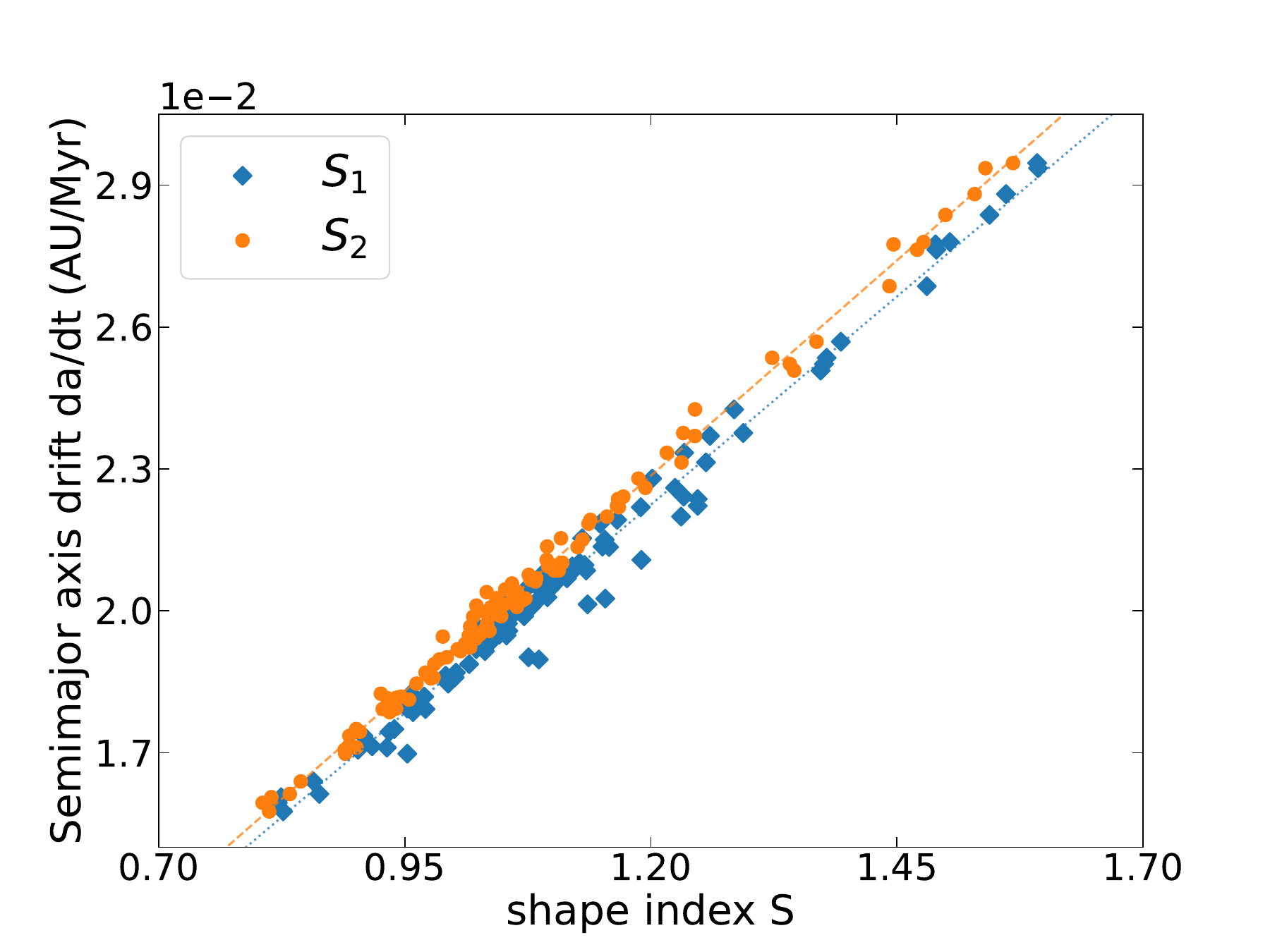}
		\put(2,70){(a)}
		\label{a}
	\end{overpic}
	\hspace{0mm}
	\begin{overpic}[width = 9cm]{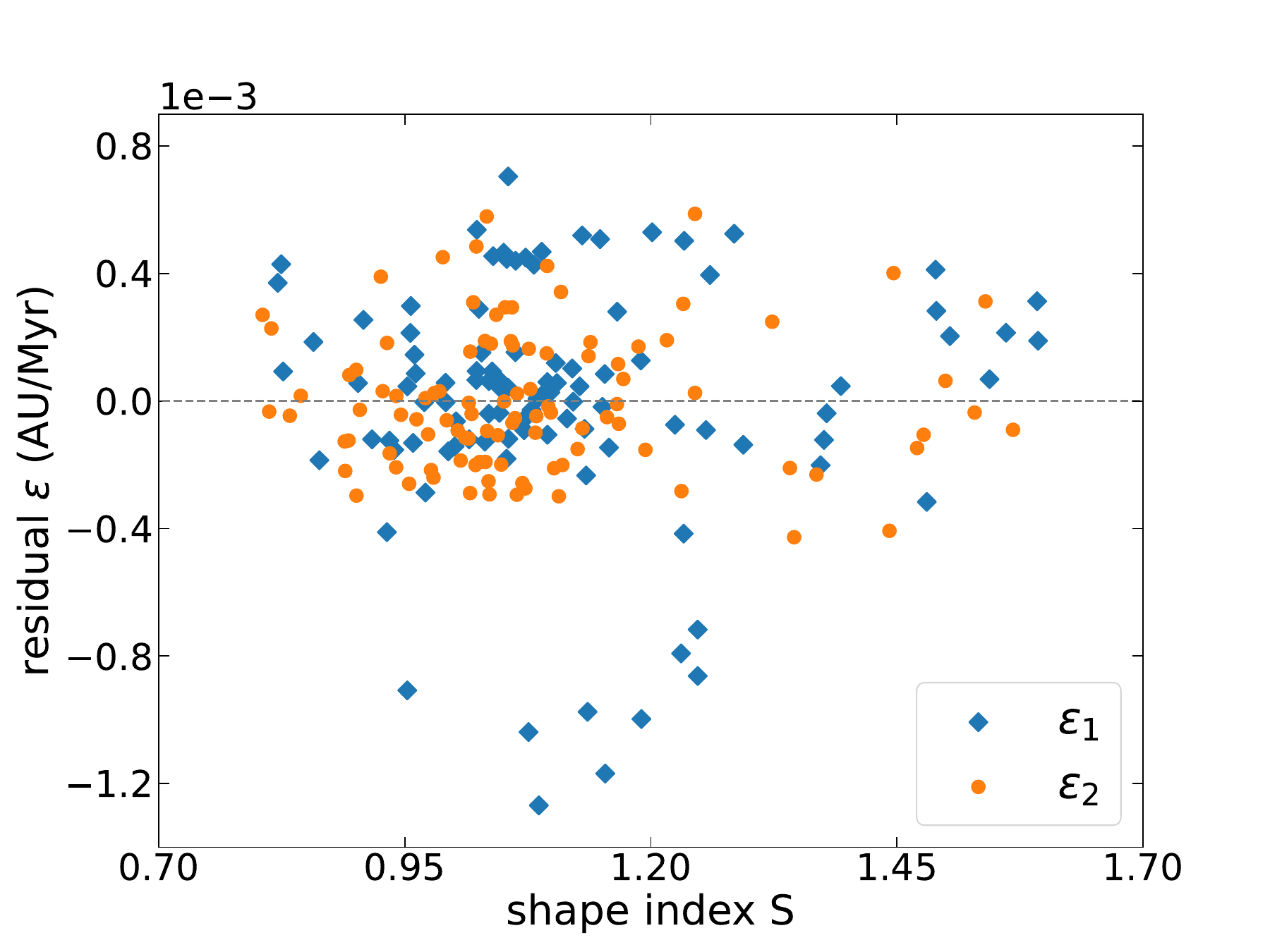}
		\put(2,70){(b)}
	\end{overpic}
	
	\begin{overpic}[width = 4.3cm]{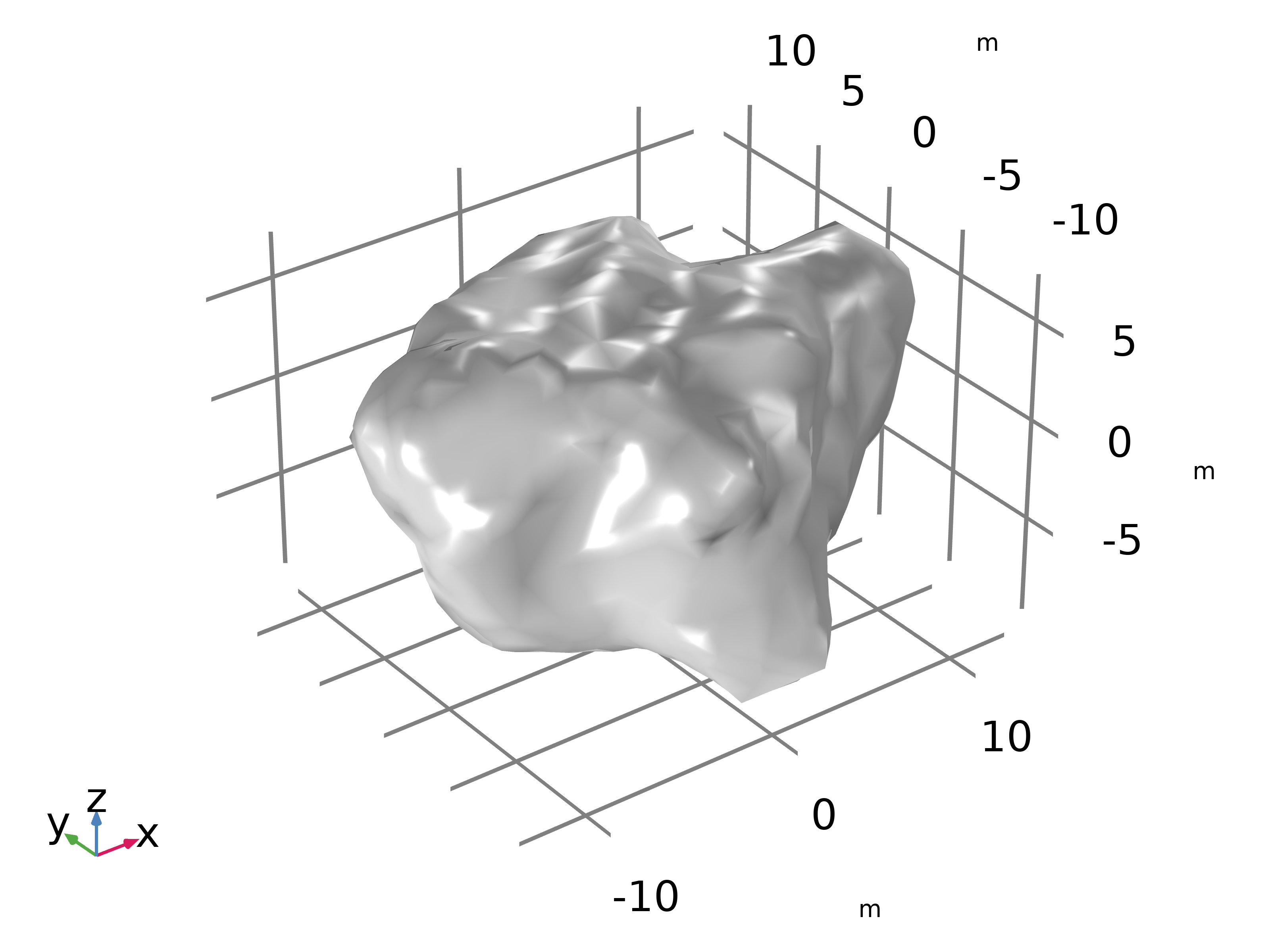}
		\put(2,70){(c)}
	\end{overpic}
	\hspace{0mm}
	\begin{overpic}[width = 4.3cm]{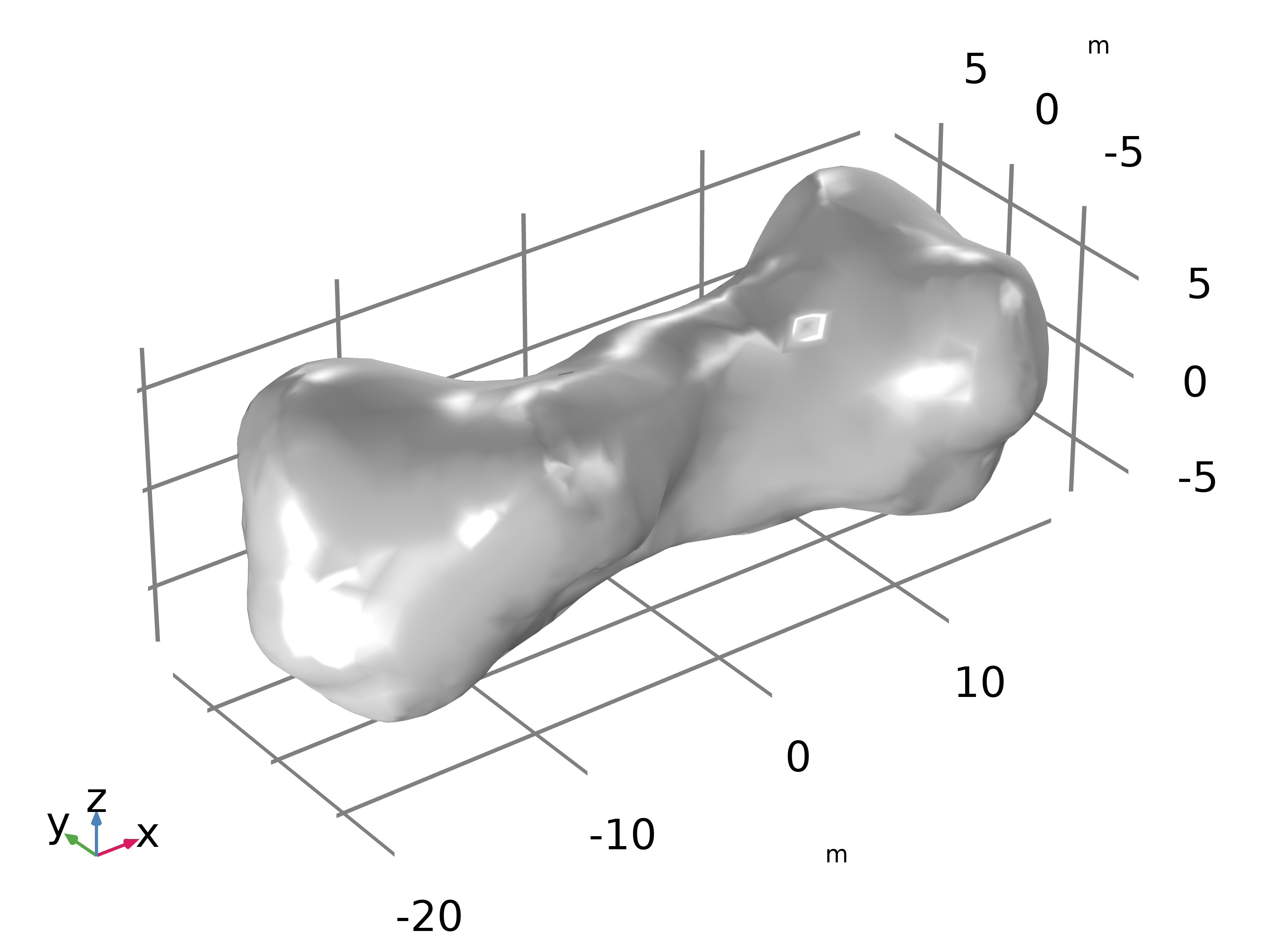}
	\end{overpic}
	\hspace{0mm}
	\begin{overpic}[width = 4.3cm]{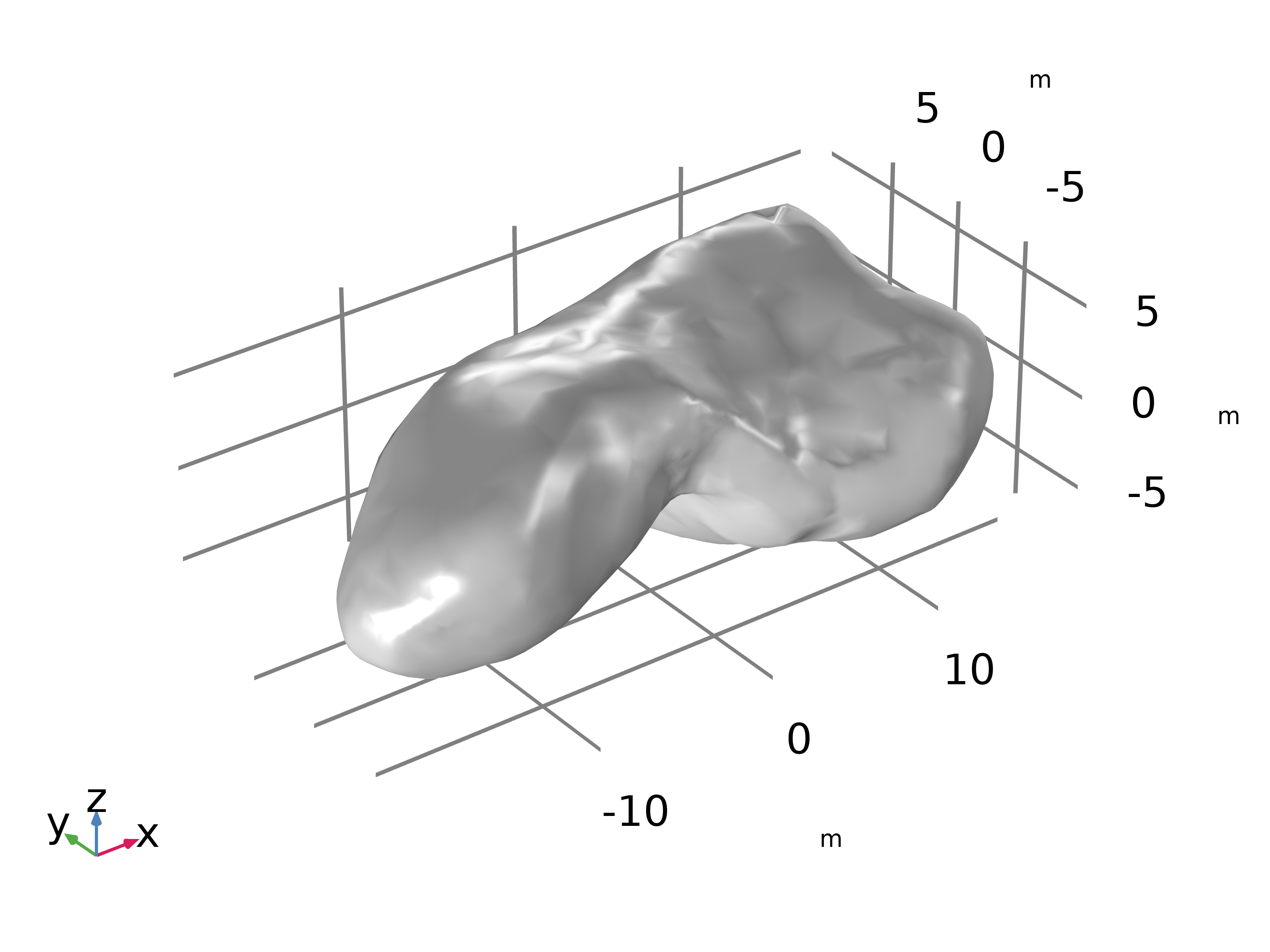}
	\end{overpic}
	\hspace{0mm}
	\begin{overpic}[width = 4.3cm]{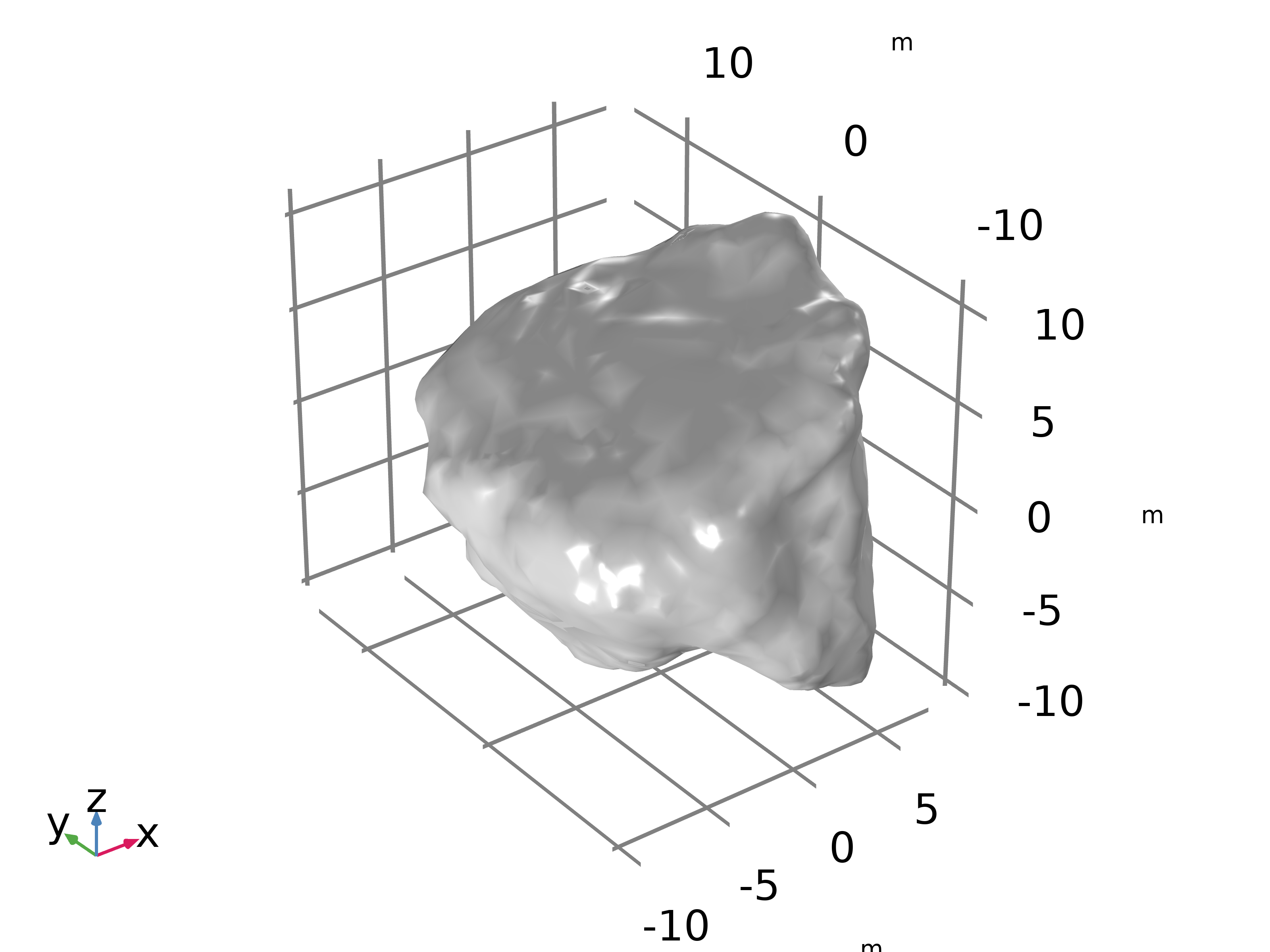}
	\end{overpic}
	\caption{Relationship between the semi-major axis drift rate ($\dif a/\dif t$) and the shape index ($S_{1,2}$). (a) $\dif a/\dif t$ versus $S_1$ (cyan diamonds) and $S_2$ (orange dots). The $\dif a/\dif t$ values are derived from numerical simulations by COMSOL (see text). The dotted and dashed lines are the least squares fitting for $S_1$ and $S_2$, respectively. (b) Fitting residuals $\varepsilon$ for $S_1$ (cyan diamonds) and $S_2$ (orange dots). The grey dashed line is $\varepsilon = 0$. (c) Four shape models with large residuals in the fitting of $\dif a/\dif t$ versus $S_1$. From left to right, they have the shapes as asteroids YORP, Kleopatra, Ida, and Golevka.}
	\label{fig:S1S2}
\end{figure*}

We present in Fig.~\ref{fig:S1S2}(a) the semi-major axis drift rates ($\dif a/\dif t$) versus the two shape indices ($S_1$, $S_2$) for these 102 shape models. The linear fits to the relationship between $\dif a/\dif t$ and $S_{1,2}$ are also plotted in Fig.~\ref{fig:S1S2}(a), and the residuals of the fits are shown in Fig.~\ref{fig:S1S2}(b). 

Just as in \citet{xu2022diurnal}, the linear dependence of $\dif a/\dif t$ on the shape index $S_1$  (cyan diamonds in Fig.~\ref{fig:S1S2}(a)) can be easily recognised. The linear fit has a fairly impressive coefficient of determination of $R^2 = 0.984$ and the mean standard deviation is $\sigma_{\varepsilon 1}=3.8\times 10^{-4}$\,AU/Myr. However, the distribution of residuals (cyan diamonds in Fig.~\ref{fig:S1S2}(b)) exhibits anomalies, and some significant outliers, all with negative values, can be seen in the lower part of the panel. 

We suspect that the obvious deviation from the linear relation is due to the shadowing effect that has been neglected in $S_1$. In fact, when a facet on the asteroid surface is shadowed by other facets during rotation, it is unable to absorb energy from the solar radiation, and accordingly, its contribution to the diurnal Yarkovsky effect is weakened. For an asteroid that deviates significantly from fully convex shape, ignoring the effect of projected shadow may lead to a significant overestimation of its shape index $S_1$. An overestimated $S_1$ corresponds to a biased high semi-major axis drift in the fit, thus results in a negative residual. We examine the shape models that are clearly outliers in Fig.~\ref{fig:S1S2}(b)  (blue diamonds for $S_1$), and find that they are really affected by the shadowing effect. As examples, we present four shape models with the most anomalous residuals in Fig.~\ref{fig:S1S2}(c), all of which, as expected, are extremely irregular shapes with significant projected shadows. 

Taking the shadowing effect into account, the shape index $S_2$ behaves better in indicating the strength of Yarkovsky effect than $S_1$, demonstrating a more distinct linear relationship with $\dif a/\dif t$ (orange dots in Fig.~\ref{fig:S1S2}(a)). All the points, including those outliers in the fitting results of $S_1$, now fit closely the line. The coefficient of determination of the linear fit now is $R^2=0.995$, and the corresponding residuals in Fig.~\ref{fig:S1S2}(b) form a normal distribution, with the mean standard deviation decreasing to $\sigma_{\varepsilon 2}=2.2\times 10^{-4}$\,AU/Myr.

\subsection{Dependence on K}

\label{sec:deponk}

When defining the shape index, the absorption, conduction and radiation of energy via a certain surface element are assumed to occur only in the same surface element, that is, the lateral heat transfer has been neglected. When establishing the linear relationship (Fig.~\ref{fig:S1S2}) between the diurnal Yarkovsky effect and the shape indices, a thermal conductivity $K_1=0.0015$\,Wm$^{-1}$K$^{-1}$ was chosen, under which the heat diffusion should be so slow that the assumption of null lateral heat transfer between surface elements is fairly reasonable. However, for basalt or iron-rich asteroids with much higher thermal conductivity, the lateral heat transfer cannot be ignored and the temperature distribution on the surface may be significantly influenced.

For large ($R\gg l_{\rm d}$) asteroids, the Yarkovsky effect is basically determined by the thermal parameter $\Theta$. In the above calculations, the  $K_1=0.0015$\,Wm$^{-1}$K$^{-1}$ with other parameters given in Table~\ref{tab:para} corresponds to a relatively small thermal parameter $\Theta_1 = 0.736$, although $\Theta$ is typically of the order of unity \citep{2015aste.book..509V} for objects found near Earth or in the main belt. The bulk density $\rho$ and heat capacity $C$ of asteroids may vary over a range of multiples, while the value of conductivity $K$ has a much greater range of more than four orders of magnitude \citep{delbo2015asteroid}. To investigate how the heat conduction may influence the relationship between $\dif a/\dif t$ and $S_2$, using the same shape models and the same $\rho$ and $C$ values as in Fig.~\ref{fig:S1S2}, we calculated the Yarkovsky effect of asteroids with two additional conductivity values $K_2=0.15$ and $K_3=15$\,Wm$^{-1}$K$^{-1}$, corresponding to thermal parameter $\Theta_2 = 7.36$ and $\Theta_3 = 73.6$, and check whether the linear relationship between the strength of Yarkovsky effect and the shape index can be verified. We note that these parameters adopted here including the $\Theta$ may be not typical values for real asteroids. But our goal is to verify the applicability of the shape index, even under very extreme conditions. In this sense, these parameters are acceptable. 

\begin{figure}[!htbp]
\centering
\begin{overpic}[width=9cm]{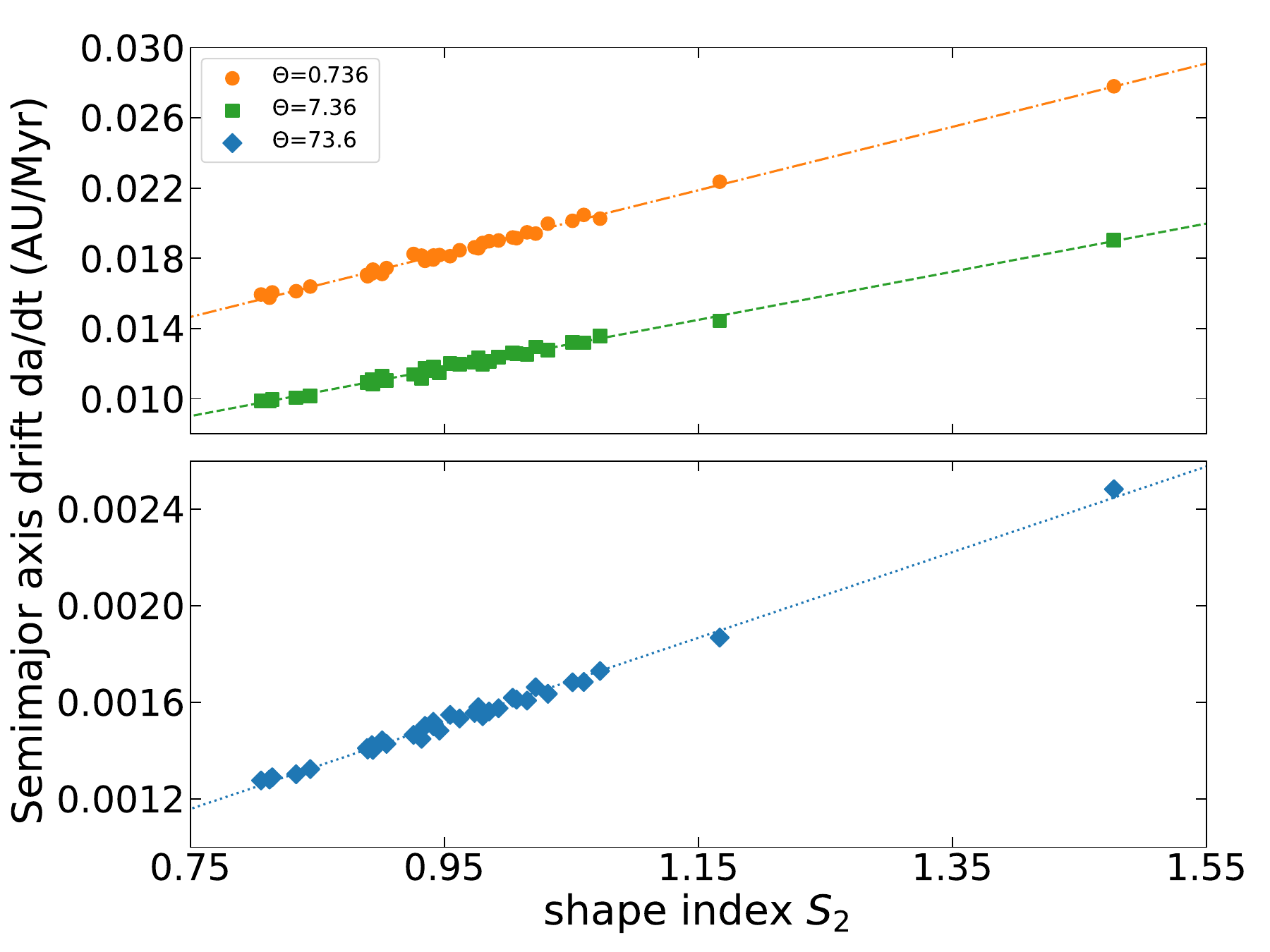}
\end{overpic}
\caption{Semi-major axis drift rate $\dif a/\dif t$ for different thermal parameter $\Theta$ versus shape index $S_2$. The cyan dots, green squares and cyan diamonds represent the $\dif a/\dif t$ for asteroids with different $K$ values, thus of $\Theta_1=0.736$, $\Theta_2=7.36$ and $\Theta_3=73.6$, respectively. The lines are the corresponding least-squares fits.}
\label{fig:S2VarTheta}
\end{figure}

To reach the dynamic equilibrium of temperature distribution on the surface, 100 and 150 rotation periods are required in the numerical simulations for the cases of $K_2$ and $K_3$, respectively. Limited by our computational resource, this time we calculated only 34 asteroid models spinning all around the $z$-axis in these simulations. The results, including those for $K_1$, are summarised in Fig.~\ref{fig:S2VarTheta}. 

As $K$ increases by orders of magnitude, $\Theta$ deviates away from 1, and for a given shape (thus a given $S_2$), the semi-major axis drift rate decreases. For example, at $S_2=0.95$, we obtain from Fig.~\ref{fig:S2VarTheta} that the $\dif a/\dif t$ is $0.01827, 0.01176$ and $0.001513$\,AU/Myr for $\Theta_1, \Theta_2$ and $\Theta_3$, respectively. 
In fact, a relative larger $\Theta$ ($>1$) indicates that the energy is being efficiently stored in the surface layer and the temperature is more likely to be evenly distributed on the surface, resulting in a weaker Yarkovsky effect.
However, as shown in Fig.~\ref{fig:S2VarTheta}, the linear relationship between $\dif a/\dif t$ and $S_2$ holds very well for all three $\Theta$ values. The least-squares fitting gives the relations as:
\begin{equation} \label{eq:fitline}
\frac{\dif a}{\dif t}=\left\{
\begin{aligned}
 0.01804 S_2 &+ 0.001130  & \text{ for } &\Theta_1, \\
 0.01371 S_2 &- 0.001264 & \text{ for } &\Theta_2, \\
 0.001773 S_2 &- 0.0001709 & \text{ for }&\Theta_3. 
\end{aligned}
\right.
\end{equation}
Note that all units (AU/Myr) have been omitted for simplicity in the equations. The coefficient of determination $R^2$ and mean standard deviation $\sigma_{\varepsilon}$ are 0.992 and $1.5\times 10^{-4}$\,AU/Myr for $\Theta_2$. For the case of $\Theta_3$, they are $R^2 = 0.993$ and $\sigma_{\varepsilon}=1.8 \times 10^{-5}$\,AU/Myr, respectively.
In addition, comparing the $\dif a/\dif t$ calculated using Eq.\,\eqref{eq:fitline} with the real values obtained from COMSOL simulations, we find that the maximum deviation is always less than 3\% for all these 34 models.

\begin{figure}[!htbp]
	\centering
	\begin{overpic}[width=9cm]{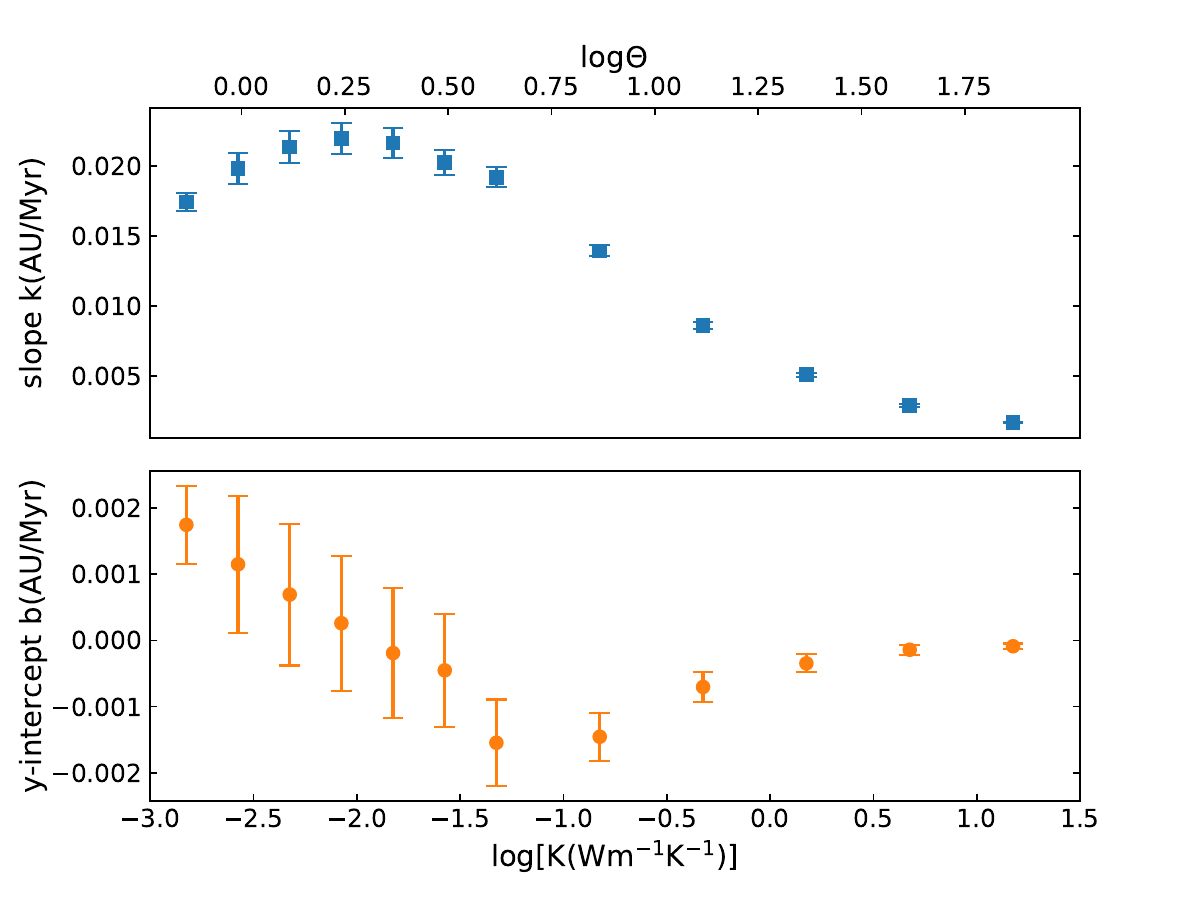}
	\end{overpic}
	\caption{Parameters ($k$ and $b$ in Eq.\,\eqref{eq:linear}) of linear fitting for different conductivity $K$ (and thus thermal parameter values $\Theta$). Note the errors for large $K$ values are so small that the error bars are narrower than the symbols standing for data points. }
	\label{fig:knb}
\end{figure}

Encouraged by these very good $R^2$ values and small deviations $\sigma_{\varepsilon}$ for cases with very different $K$ (thus different $\Theta$) values, we calculate the Yarkovsky effect for more cases with $K$ in a wide range, and fit the results as the linear relation:
\begin{equation} \label{eq:linear}
\dif a/\dif t=k S_2 + b.
\end{equation}
To save computations, for each $K$ value we selected from the 34 samples arbitrarily 5 models (with shape indices $S_2\approx 0.805, 0.833, 0.962, 0.993, 1.05$), calculated their surface temperature using COMSOL, derived the $\dif a/\dif t$, and finally fitted the results as Eq.~\eqref{eq:linear}. The slope $k$ and the intercept $b$ are summarized in Fig.~\ref{fig:knb}. 
We note that the $k$ and $b$ for cases of $K_1, K_2$ and $K_3$ (thus $\Theta_1, \Theta_2$ and $\Theta_3$) in Fig.~\ref{fig:knb} were re-computed from the same 5 shape models. 

Ideally, if the Yarkovsky effect depends on the shape of an asteroid characterised by the shape index, we would expect an intercept $b=0$ in Eq.\,\eqref{eq:linear}. But some approximations have been unavoidably introduced in the definition of shape index, for instance, as shown in Fig.~\ref{fig:angle}, the delay angles of different surface elements on a body are not strictly identical. In addition, a 3D model is used in calculating the Yarkovsky effect, but the lateral heat conduction is ignored in the definition of shape index. However, as shown in Fig.~\ref{fig:S2VarTheta}, a linear relationship is very well obeyed for shape models with the shape index $S_2$ from 0.8 to 1.5. And, as we can see in Fig.~\ref{fig:knb} and Eq.\,\eqref{eq:fitline}, the intercept $b$ is at least one order of magnitude smaller than the slope $k$, which is the exact $\dif a/\dif t$ at $S_2=1$. Therefore, for practical use, the linear relationship is still an excellent approximation.

In Fig.~\ref{fig:knb}, as the conductivity (thus $\Theta$) increases, the slope $k$ of the linear fit first increases and then decreases, reaching its maximum when $\Theta\sim 1.74$. In fact, $\Theta$ is the ratio of the thermal relaxation time to the rotational period \citep{Farinella1998}. Theoretically, for $\Theta<1$, energy is rapidly lost by radiation, and the delay angle is small, while for $\Theta>1$, the temperature gradient on the surface can not be efficiently established. Both lead to a smaller Yarkovsky effect.

When the slope $k$ reaches its maximum, we notice the intercept $b$ approaches to zero (Fig.~\ref{fig:knb}), implying the optimal condition for the definition of shape index is attained. 

\subsection{Lateral heat conduction}
\label{sec:lateral}
\begin{figure*}[!htb]
\centering
\begin{overpic}[width = 5.5cm]{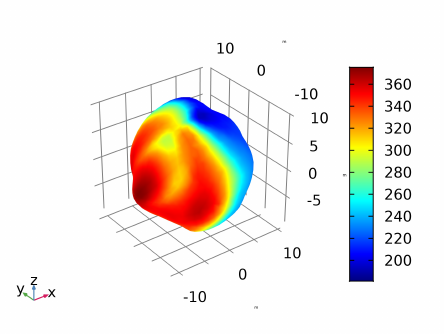}
	\put(5,65){$\Theta=0.736$}
\end{overpic}
\hspace{0mm}
\begin{overpic}[width = 5.5cm]{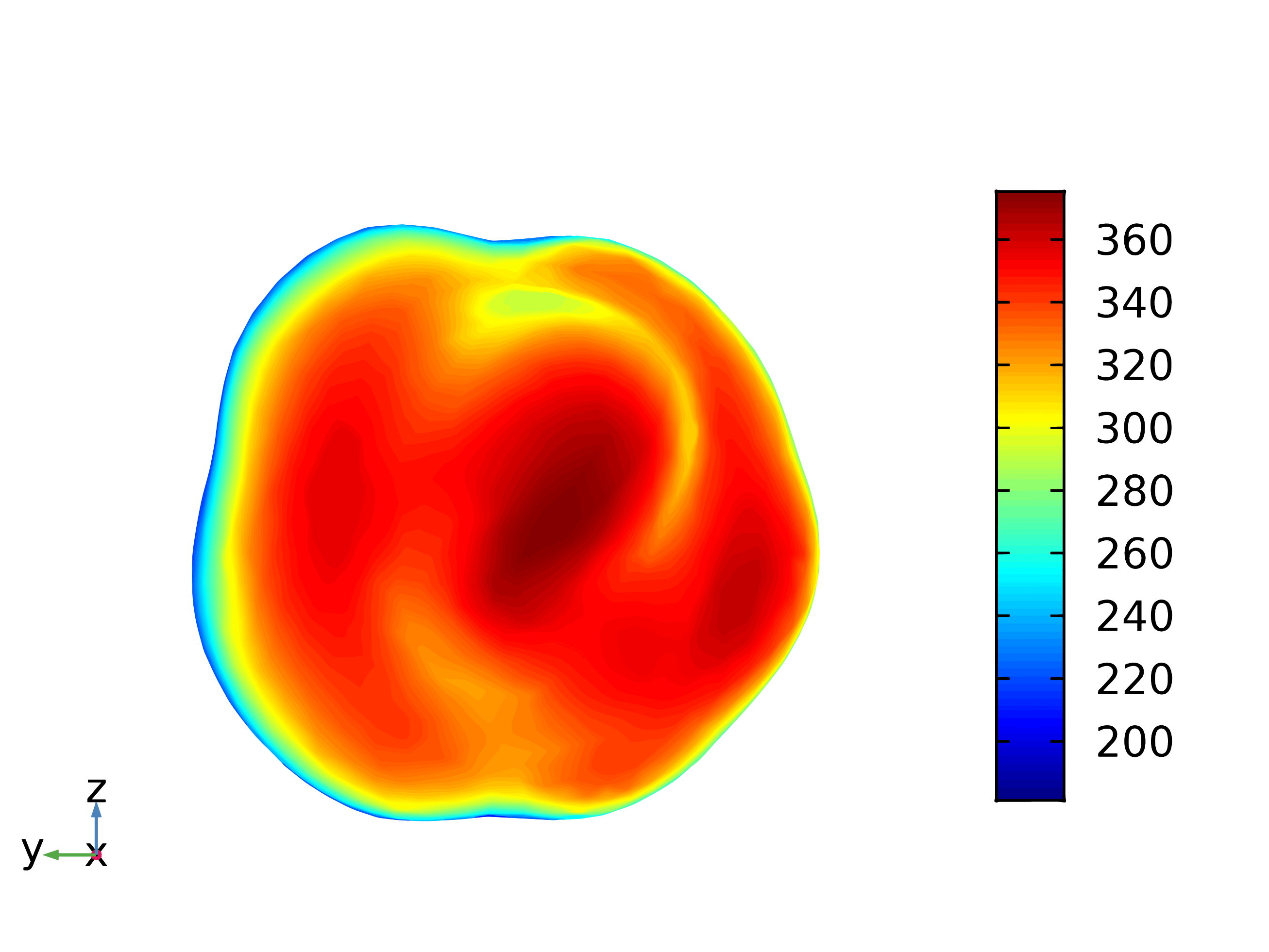}
\end{overpic}
\hspace{0mm}
\begin{overpic}[width = 5.5cm]{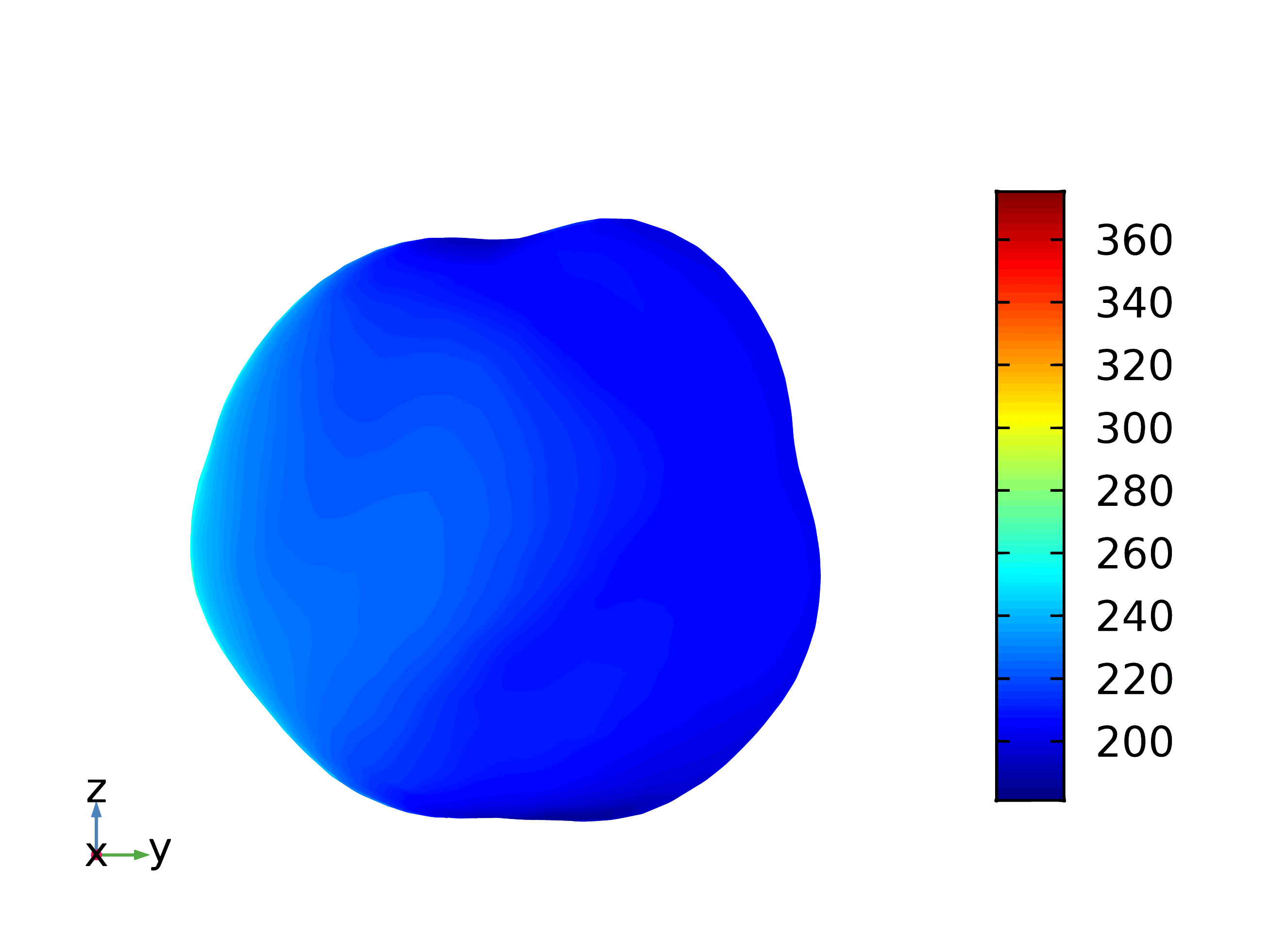}
\end{overpic}
\hspace{0mm}
\begin{overpic}[width = 5.5cm]{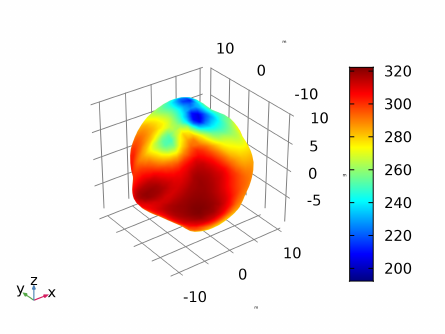}
	\put(5,65){$\Theta=7.36$}
\end{overpic}
\hspace{0mm}
\begin{overpic}[width = 5.5cm]{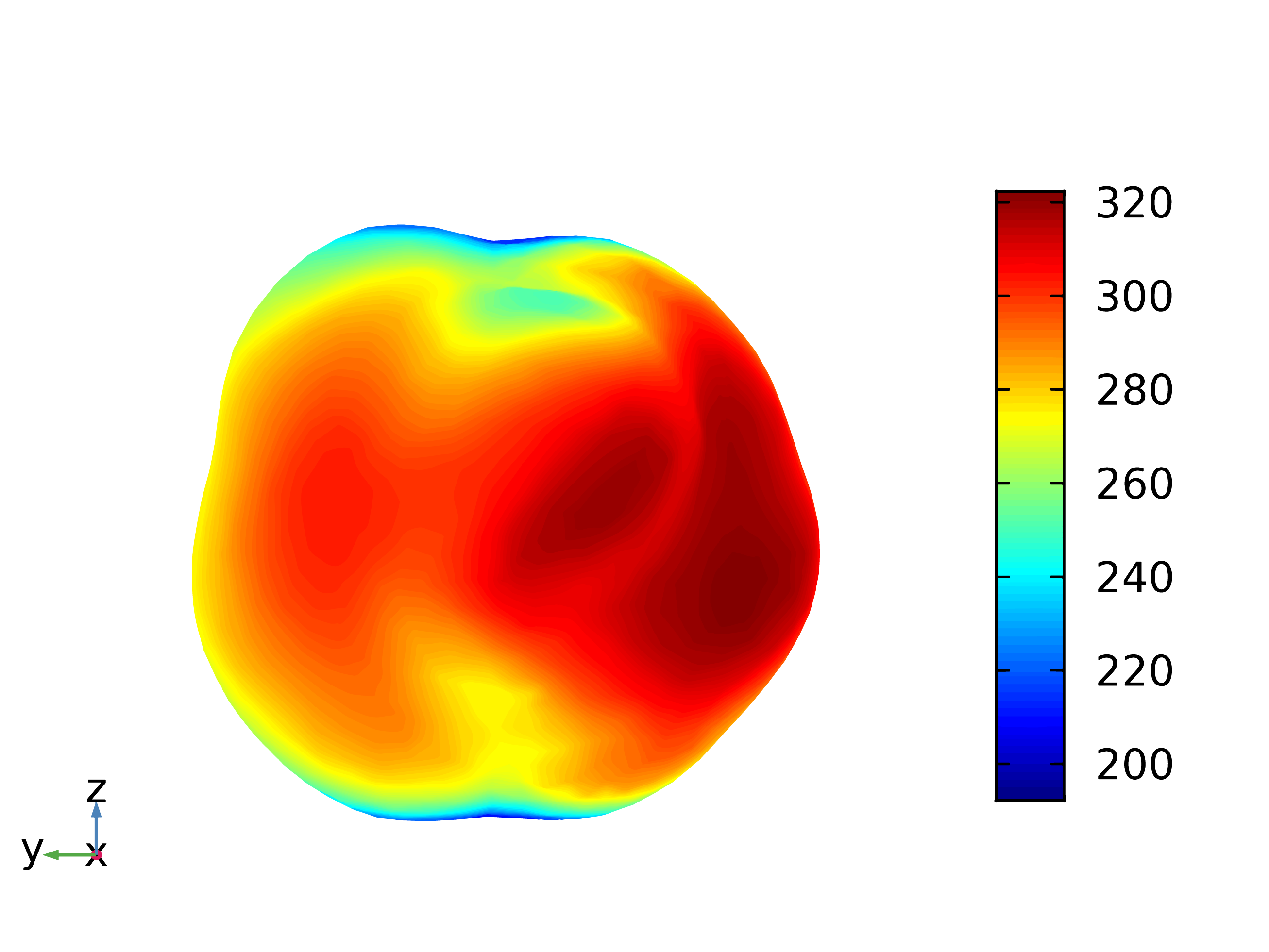}
\end{overpic}
\hspace{0mm}
\begin{overpic}[width = 5.5cm]{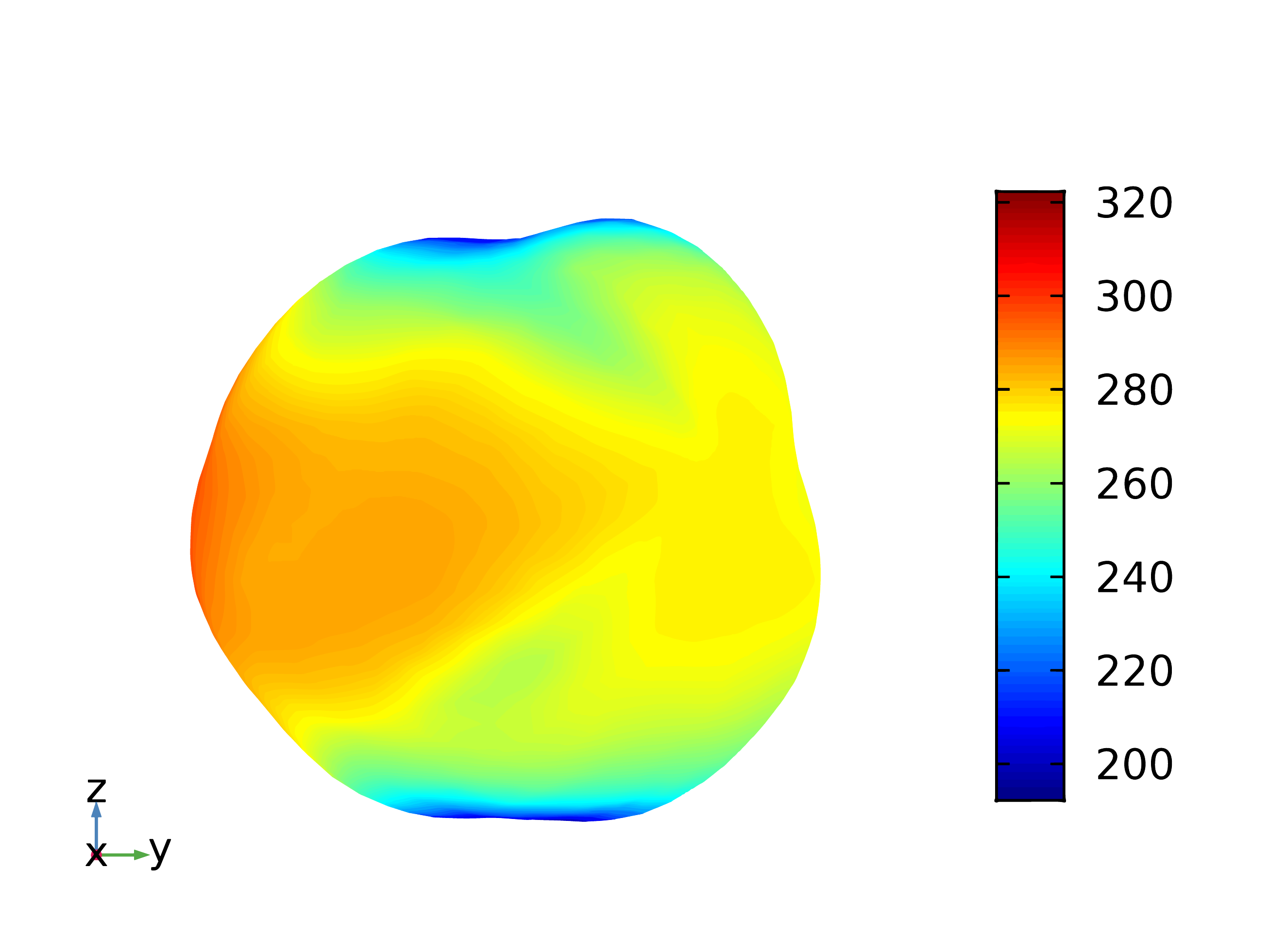}
\end{overpic}
\hspace{0mm}
\begin{overpic}[width = 5.5cm]{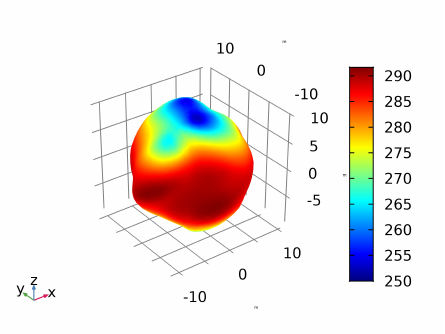}
	\put(5,65){$\Theta=73.6$}
\end{overpic}
\hspace{0mm}
\begin{overpic}[width = 5.5cm]{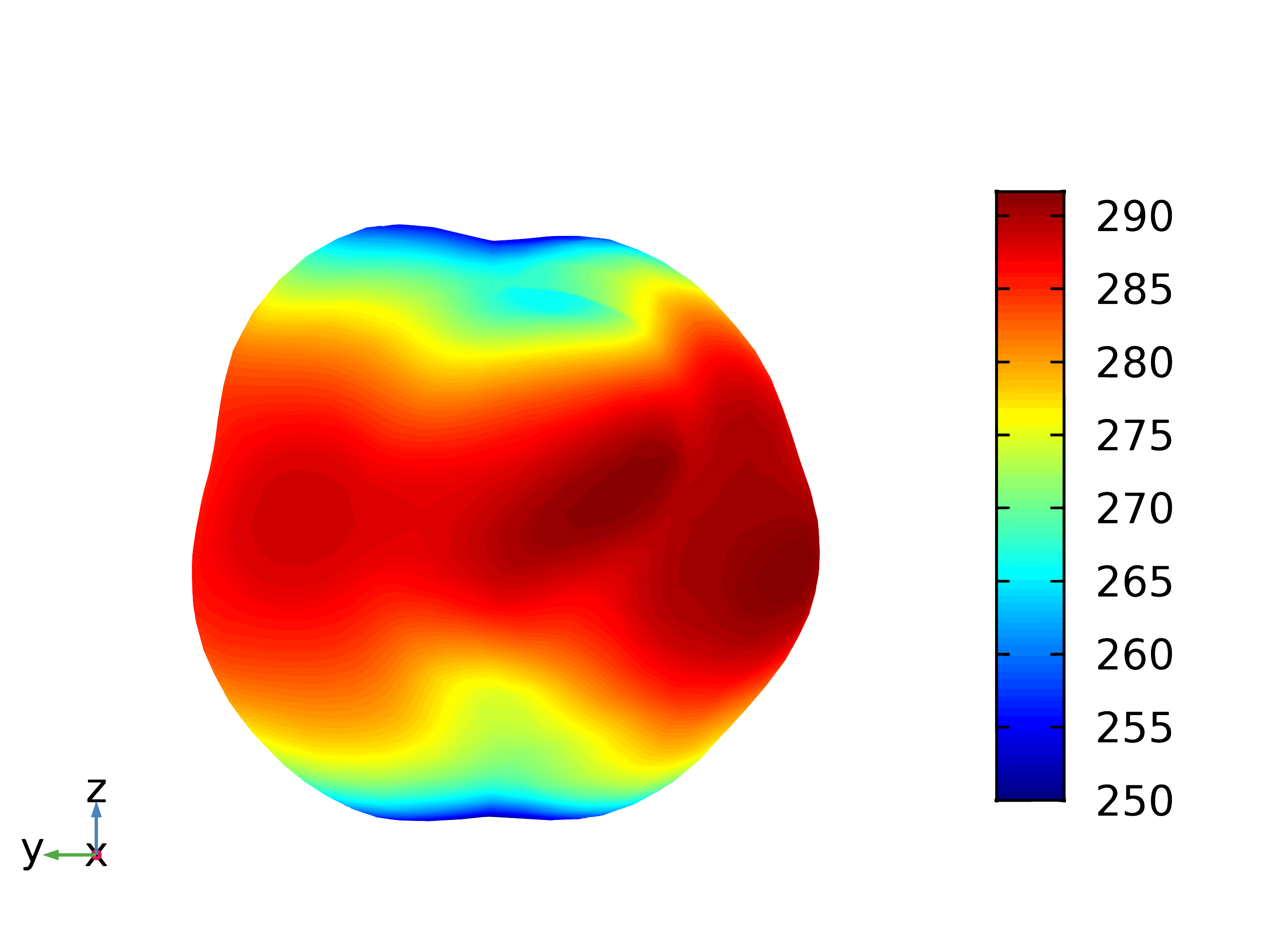}
\end{overpic}
\hspace{0mm}
\begin{overpic}[width = 5.5cm]{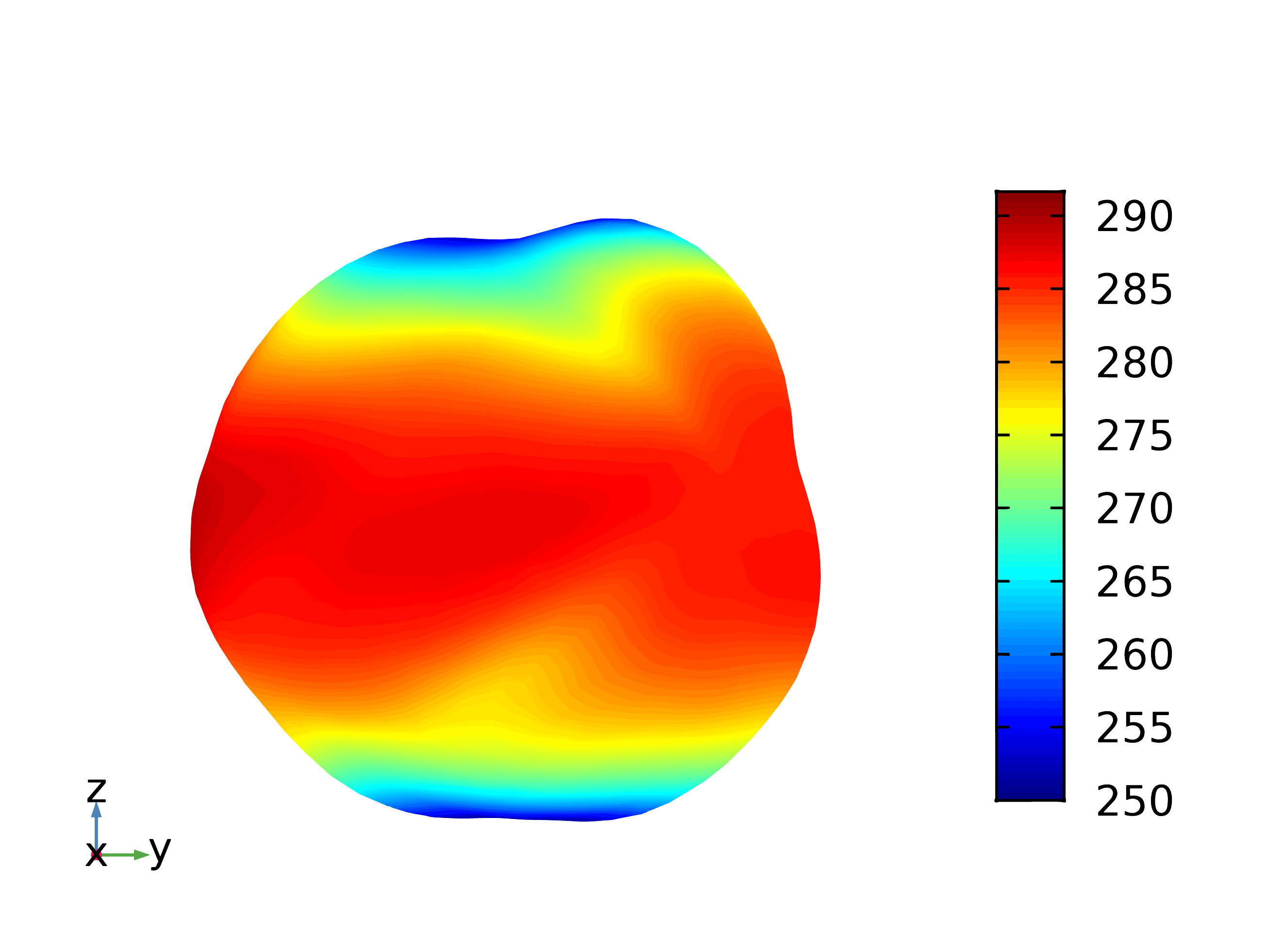}
\end{overpic}
\hspace{0mm}
\caption{Temperature distribution on the surface of asteroids with different thermal parameter $\Theta$, taking the shape of 1998 KY26 as an example. The small body has a retrograde rotation around the $z$-axis, with the solar radiation from the direction $\textbf{n}_{\rm in}=\left(-1,0,0\right)$. From top to bottom are the cases for $\Theta_1, \Theta_2$ and $\Theta_3$, and from left to right are the isometric views, day hemispheres and night hemispheres. Note that the colour bars have different temperature scales. }
\label{fig:tempdis}
\end{figure*}

The rate of heat conduction inside a body is determined not only by the value of $K$, but also by the temperature gradient. A higher $\Theta$ leads to a lower temperature difference between adjacent surface elements. Thus, as we have shown in Fig.~\ref{fig:S2VarTheta} and Fig.~\ref{fig:knb}, a larger $\Theta$ value does not necessarily fail the mechanism of the linear relationship between $\dif a/\dif t$ and $S_2$. We take the asteroid 1998 KY26 as a shape example to show the variation of temperature distribution on the surface at different $\Theta$ values. Again, three $K$ values (thus three thermal parameters $\Theta_1, \Theta_2, \Theta_3$) are adopted and the results are presented in Fig.~\ref{fig:tempdis}. 

The small $K$ value (corresponding to $\Theta_1$, the top panels of Fig.~\ref{fig:tempdis}) prohibits the diffusion of energy in the body, and the surface elements around the equator absorb and keep much more solar radiation than the ones in the polar region. As a result, the global temperature difference across the surface reaches nearly 200\,K, that is, the temperature gradient on the surface is large. But the thermal conductivity is so small that the lateral heat transfer is nearly blocked. Thus the assumption of null lateral heat transfer in the definition of shape index is fulfilled, and the linear dependence of Yarkovsky effect on the shape index is soundly obeyed. 

When $K$ increases by 4 orders of magnitude to $K_3$ (thus $\Theta$ increases by 2 orders of magnitude to $\Theta_3$) the global temperature difference across the surface is only $\sim$40\,K (bottom panels of Fig.~\ref{fig:tempdis}). The small temperature gradient limits the lateral heat transfer, and therefore, the vast majority of heat received by a surface element is in fact stored in and then emitted after a delay from the same surface element. The assumption of negligible lateral diffusion of heat is still available, and the linear relationship between the semi-major axis drift and $S_2$ still holds. 

\begin{figure}
	\centering
	\begin{overpic}[width=9cm]{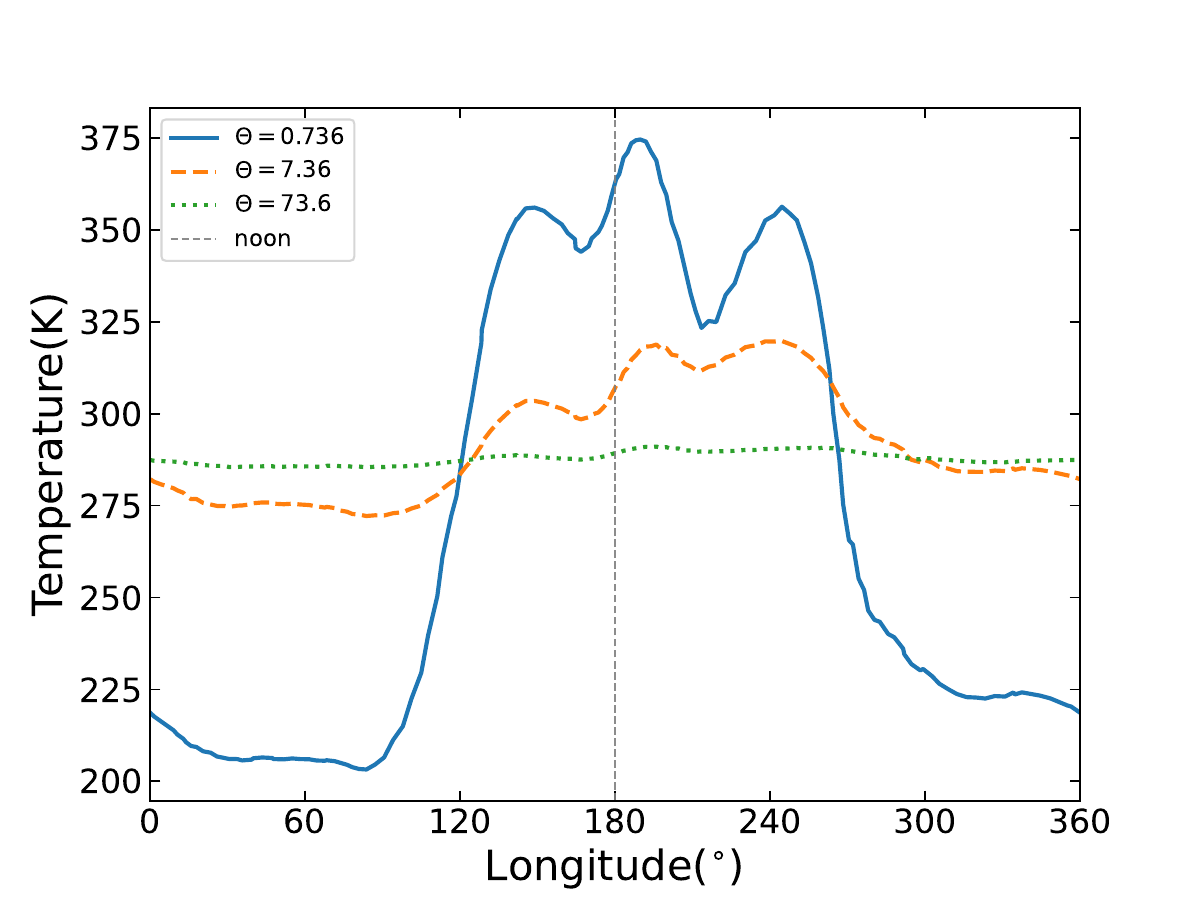}
	\end{overpic}
	\caption{Surface temperature on equator of the asteroid model in Fig.~\ref{fig:tempdis}. The longitude of $0^\circ$ (midnight) and $180^\circ$ (noon) are on the meridian plane. Three $K$ values (corresponding to $\Theta_1, \Theta_2, \Theta_3$) are adopted and the results are given in different colours.}
	\label{fig:equatemp}
\end{figure}

Comparing the surface temperatures in the three cases in Fig.~\ref{fig:tempdis}, we find that the most obvious change due to different $\Theta$ values is the difference between the temperatures of day side and night side. For any given $\Theta$ value, the greatest temperature difference on the surface is always between the equatorial and polar regions. As the thermal conductivity increases, the magnitude of temperature variation along the latitude circles (as an example, the temperature along the equator is plotted in Fig.~\ref{fig:equatemp}) decreases much more than it does along the longitude circles. This again indicates that the lateral heat transfer contributes little to the temperature distribution, otherwise the polar region should obtain more energy from lateral diffusion of heat to reach a higher temperature and a lower temperature variation along the longitude circle. In addition, as $\Theta$ increases, the temperature distribution on the surface, particularly along the east-west direction, maintains nearly the same pattern in the three cases (Fig.~\ref{fig:tempdis}), implying also that the heat diffusion is enhanced mainly in the direction into the deeper part inside the asteroid, which leads to a longer delay of thermal emission and a lower temperature fluctuation on the surface. 

Finally, the influence of the lateral heat transfer can also be estimated by comparing the Yarkovsky effects of ideal spherical asteroid calculated using both 1D and 3D models. In the 1D model, the lateral heat transfer is completely neglected, while in the 3D model it is fully taken into account. Regarding the Yarkovsky effect ($\dif a/\dif t$) calculated from the 3D model as accurate, we find the relative errors of the 1D model are $1.6\times 10^{-4}$, $3.2\times 10^{-4}$ and $2.3\times 10^{-3}$ when the conductivity is set as $K_1, K_2$ and $K_3$, respectively. Although the relative error (due to neglecting the lateral heat transfer) increases somewhat as the thermal conductivity greatly increases (by four orders of magnitude), it remains at a low level of $10^{-3}$. This suggests that the lateral heat transfer does not influence the Yarkovsky effect much, and a high $K$ will not invalidate the linear relationship between the semi-major axis drift rate ($\dif a/\dif t$) and the shape index ($S_2$), as long as the size of asteroid is not too small ($R \gg l_{\rm d}$).

\subsection{Scattering effect and self-heating effect}
\label{sec:selfheat}

In addition to absorbing the Solar radiation, a surface element may also obtain energy from other elements by their scattering (the solar radiation) and thermal irradiation (this effect is called self-heating). This part of energy is ignored in the definition of shape index $S_2$. 

The contribution from the scattering and self-heating effects can be numerically estimated. We select the four shape models in Fig.\ref{fig:S1S2}(c) as examples, since they deviate significantly from fully convex shapes and the influences of scattering and self-heating effects are expected to be most evident in these models. The `real' Yarkovsky effects of these four models have been calculated and presented in Fig.\ref{fig:S2VarTheta}, where the scattering and self-heating effects were taken into account. We regard the tangential component of the recoil force ($F_t$ in Eq.\eqref{eq4}) as the accurate measurement of the Yarkovsky effect, and numerically re-calculate the force component $F'_t$ for the same four models but without the scattering and self-heating effects (by turning off these effects in the COMSOL simulations). The relative differences between them ($\left|1-F'_t/F_t\right|$) are summarized in Table~\ref{tab:self-heat}. 

\begin{table}
	\centering
	\caption{Relative differences ($\left|1-F'_t/F_t\right|$) arising from the scattering and self-heating effects for four shape models (see text) with three thermal conductivity $K$ values. When $K_1$ is adopted, the differences between $F'_t$ and $F_t$ are much smaller than the control accuracy in our simulations, therefore we just list the upper limit. }
	\begin{tabular}{c | c c c c}
		\hline
		\hline
		~  & YORP & Kleopatra & Ida & Golevka \\
		\hline
		$K_1$ & $\le 10^{-5}$ & $\le 10^{-5}$ & $\le 10^{-5}$ & $\le 10^{-5}$ \\
		$K_2$ & $4.6\times 10^{-4}$ & $3.1\times 10^{-4}$ & $2.6\times 10^{-4}$ & $4.6\times 10^{-4}$ \\
		$K_3$ & $1.1\times 10^{-3}$ & $1.9\times 10^{-3}$ & $1.5\times 10^{-3}$ &$1.7\times 10^{-3}$ \\
		\hline
	\end{tabular} \label{tab:self-heat}
\end{table}

As shown in Table~\ref{tab:self-heat}, the relative error due to neglecting the scattering and self-heating effects increases as $K$ increases. However, even for the highest possible conductivity $K_3$ and for the most irregular shapes, the error induced by neglecting scattering and self-heating is still very small, only of the order of $10^{-3}$.

The scattering Solar radiation absorbed by the $i$-th surface element at moment $t$ can be written as
\begin{equation}
\label{eq11}
E_{\rm scat}=-\alpha\left(1-\alpha\right)\mathcal{E}\sum_{j\neq i} f_{i,j} \left(1-s_j\left(t\right)\right)\left(\textbf{n}_j\cdot\textbf{n}_{\rm in}\right).
\end{equation}
The $f_{i,j}$ in Eq.~\eqref{eq11} is the view factor from the $i$-th surface element to the $j$-th one, and it reads
\begin{equation}
\label{eq12}
f_{i,j} = v_{i,j} \frac{\cos\theta_i \cos\theta_j}{\pi d^2_{i,j}}a_j,
\end{equation}
where $v_{i,j}$ denotes whether there is line-of-sight visibility between the $i$-th and $j$-th  surface elements, $d_{i,j}$ is the distance between them, $\theta_i$ and $\theta_j$ are the incidence and emission angles of the $i$-th and $j$-th surface element, respectively, and $a_j$ is the area. For more details on the definition and calculation of view factor, see \citet{lagerros1998thermal} and \citet{rozitis2011directional}.

The self-heating effect depends on the surface temperature distribution in addition to the surface geometry of the asteroid. For the $i$-th surface element, the radiation that it absorbs from other elements is
\begin{equation}
\label{eq13}
E_{\rm rad}=\alpha \epsilon \sigma \sum_{j\neq i} f_{i,j} T^4_j\left(t\right). 
\end{equation}
The temperature $T_j$ at the $j$-th surface element can be estimated by assuming an instantaneous heat absorption and emission, that is, an equilibrium
\begin{equation}
\label{eq14}
-\alpha\mathcal{E}\left(1-s_j\left(t\right)\right)\left(\textbf{n}_j\cdot\textbf{n}_{\rm 
	in}\right)=\epsilon \sigma T^4_j\left(t\right).
\end{equation}

All the energy absorbed by the $i$-th surface element at time $t$, including contributions from the direct Solar radiation as in Eq.~\eqref{eq5} (taking into account the shadowing effect), scattering as in Eq.~\eqref{eq11} and self-heating as in Eq.~\eqref{eq13}, now is
\begin{equation}
\label{eq15}
\begin{aligned}
 E_{\rm in}+E_{\rm scat}+E_{\rm rad} = -\alpha \mathcal{E} &\Bigg[  a_i\Big(1-s_i(t)\Big)\big(\textbf{n}_i\cdot\textbf{n}_{\rm in}\big)  \\
  &+\sum_{j\neq i}f_{i,j} \Big(1-s_j(t)\Big)\big(\textbf{n}_j\cdot\textbf{n}_{\rm in}\big)\Bigg].
\end{aligned}
\end{equation}
Finally, an improved shape index $S_3$ that takes into account the scattering effect and self-heating effect can be written as:
\begin{equation} \label{eq16}
\begin{aligned}
S_3 = \frac{A}{P} \int_{0}^{P} \Bigg\{ & \sum^N_i \big(\textbf{n}_i\cdot\textbf{n}_{\rm in}\big) \Big[a_i\big(1-s_i (t)\big) \big(\textbf{n}_i\cdot\textbf{n}_{\rm in}\big) \\
& -\sum_{j\neq i} f_{i,j} \big(1-s_j (t)\big)\big(\textbf{n}_j\cdot\textbf{n}_{\rm in}\big)\Big] \Bigg\} \dif t\,.
\end{aligned}
\end{equation}

To check whether this new shape index $S_3$ improves the estimation of Yarkovsky effect, we compute the $S_3$ values for the 34 shape models and calculate the linear fitting between $\dif a/\dif t$ and $S_3$ as we did for $S_1$ and $S_2$. The coefficients of determination ($R^2$) of the fitting for both $S_2$ and $S_3$ are listed in Table~\ref{tab:R2}. 

\begin{table}
\centering
\caption{Correlation coefficients ($R^2$) of the semi-major drift rate $\dif a/\dif t$ with respect to shape indices $S_2$ and $S_3$. The value for shape index $S_1$ is available only for $K_1$ and was calculated from a sample of 102 shape models (see text in Section~\ref{sec:linear}). }
\begin{tabular}{c | c c c}
\hline
\hline
  ~  &  $R^2$ (for $S_1$) &  $R^2$ (for $S_2$) &  $R^2$ (for $S_3$) \\
\hline
$K_1$  & 0.984 & 0.995 & 0.993 \\
$K_2$  & --- & 0.992 & 0.986 \\
$K_3$  & --- & 0.993 & 0.985 \\
\hline
\end{tabular}\label{tab:R2}
\end{table}

The relatively smaller $R^2$ values for $S_3$ than the ones for $S_2$ (Table~\ref{tab:R2}) reveal that this new shape index $S_3$ does not make a better linear relationship. In fact, when we calculate the self-heating effect $E_{\rm rad}$, the surface temperature has been obtained by simply assuming an instantaneous equilibrium (Eq.\ref{eq14}). This approximation is equivalent to the assumption $K=0$, and it might introduce extra error. In fact, this explains that the index $S_3$ behaves even worse when a higher $\Theta$ value is adopted, as shown in Table~\ref{tab:R2}. Without knowing the specific temperature distribution, it is difficult to evaluate the influence of self-heating on the Yarkovsky effect. 

Considering the fact that the errors caused by the scattering and self-heating effects are ignorable in practice (Table~\ref{tab:self-heat}), we therefore abandon the idea of incorporating these  effects into the definition of the shape index. 

\subsection{Shape index and shapes}
For a given asteroid, the Yarkovsky effect is determined by its thermal parameter $\Theta$. When focusing on the shape of an asteroid with given thermal parameter, two factors may influence the Yarkovsky effect. One is the orientation of surface elements, and the other is the effect of shadowing. For the former, if the normal of a facet on the body surface is parallel to the equatorial plane, as the body rotates this facet will have the largest cross sectional area with respect both to the Solar radiation and to the tangential direction of the orbit, and it will produce the most effective Yarkovsky force. Therefore, the more a body has surface elements with normal being parallel to the equatorial plane, the stronger the Yarkovsky effect is, and according to the definition of shape index, the bigger $S_1$ and $S_2$ are. For example, a cylinder with radius $R$ and height $h$ rotating around its axis, according to Eq.\,\eqref{eq:s1}, its shape index $S_1=\left(3h/4R\right)^{1/3}$. The higher the cylinder is, the bigger $S_1$ is, and vice versa. On the contrary, if a body is oblate, most of its surface elements have normal nearly perpendicular to the equatorial plane, its shape index must be small. A very low (flat) cylinder is an extreme example. 

\begin{figure}
	\centering
	\begin{overpic}[width=9cm]{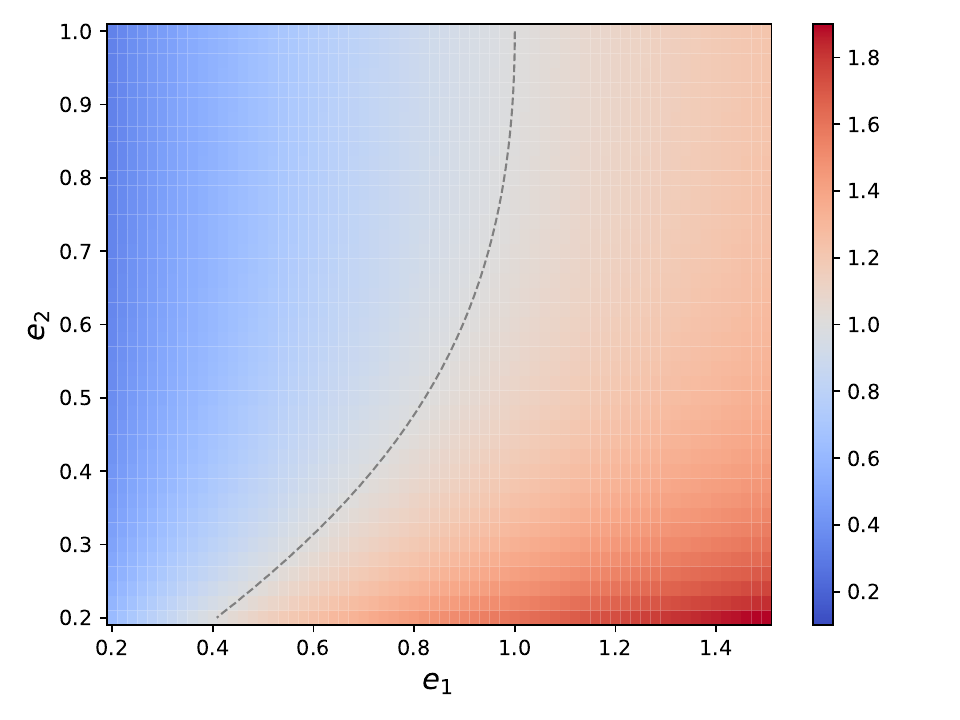}
	\end{overpic}
	\caption{The shape index $S_1$ (given by colour) for triaxial ellipsoids characterised by oblateness $e_1$ and $e_2$. The dashed line is for $S_1=1$.}
	\label{fig:ellipS1}
\end{figure}

To show more clearly the shape index value of different shapes, we consider triaxial ellipsoids with three semi-axes $\left (a, b, c\right)$. Two parameters $e_1$, $e_2$ describing the shape are defined as:
\begin{equation}
	e_1 = \frac{c}{\sqrt{ab}}, \ \ \ e_2 = \frac{b}{a}
\end{equation} 
Apparently, a triaxial ellipsoid reduces to a biaxial ellipsoid when $e_2=1$, and ellipsoids characterised by $e_2$ and $1/e_2$ are identical to each other. Assuming the ellipsoid rotates around the axis with semi-major axis $c$, we calculate the $S_1$ of them and present the results in Fig.~\ref{fig:ellipS1}. We note that ellipsoids are fully convex, so $S_1=S_2$ here.

For biaxial ellipsoids, the shape index increases as the oblateness $e_1$ increases. $S_1<1$ for oblate spheroid with $e_1<1$, and on the contrary $S_1 >1$ for $e_1>1$. For triaxial ellipsoids, the shape index $S_1>1$ for all the ellipsoids with $e_1>1$, indicating that all ``tall'' spheroids have stronger Yarkovsky effect than a sphere. While for oblate triaxial ellipsoids with $e_1<1$, the shape index can also be larger than 1, especially when $e_2$ is small. This phenomenon of oblate triaxial ellipsoids having relatively stronger Yarkovsky effect has already been observed in the numerical simulations and explained in \citet{xu2022diurnal}. It should be noted that the rotation of an ellipsoid with $e_1>1$ around the axis $c$ is generally unstable. 

\begin{figure}[!htbp]
	\centering
	\begin{overpic}[width = 4.4cm]{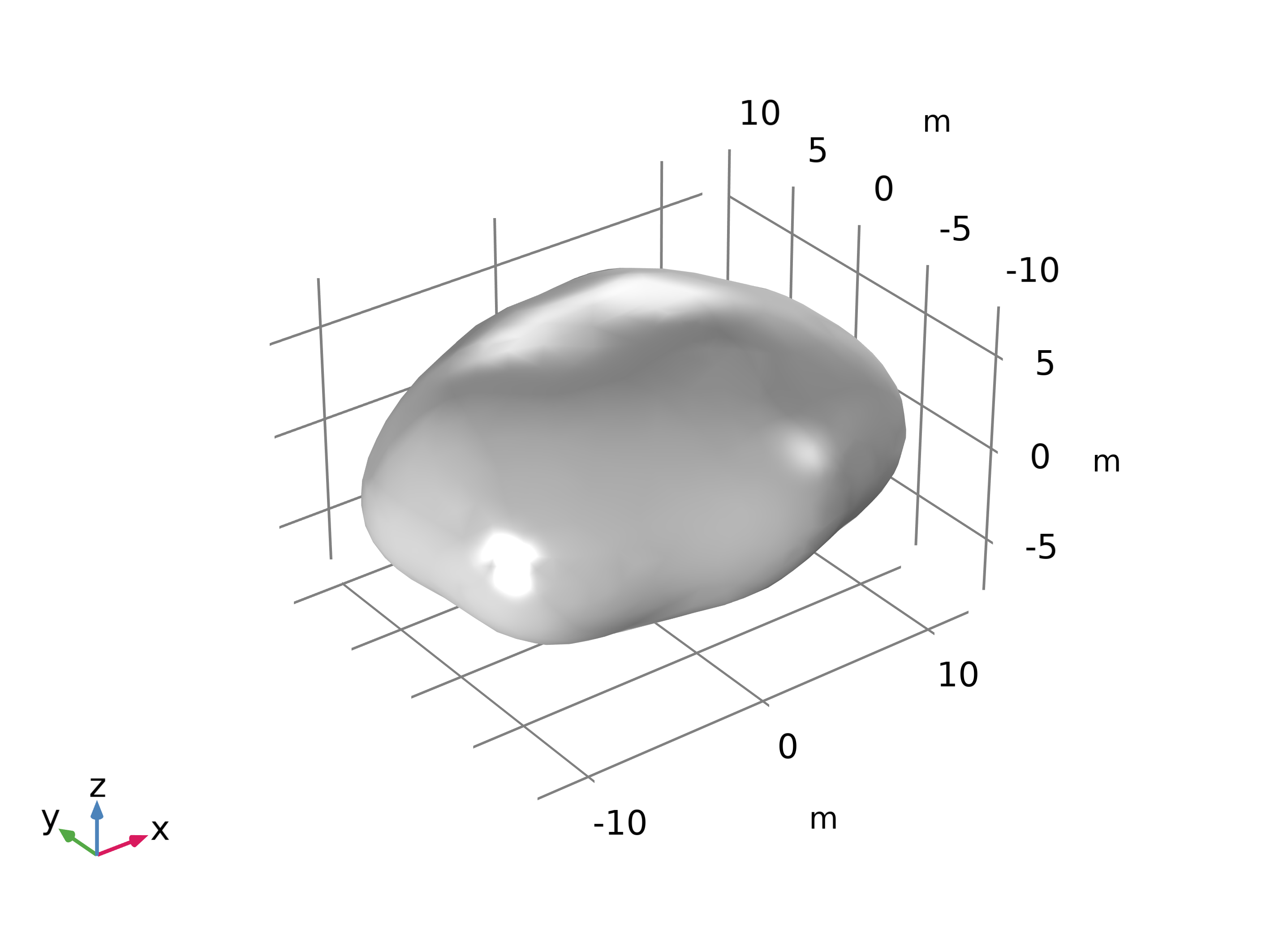}
		\put(2,70){(a)}
	\end{overpic}
	\hspace{0mm}
	\begin{overpic}[width = 4.4cm]{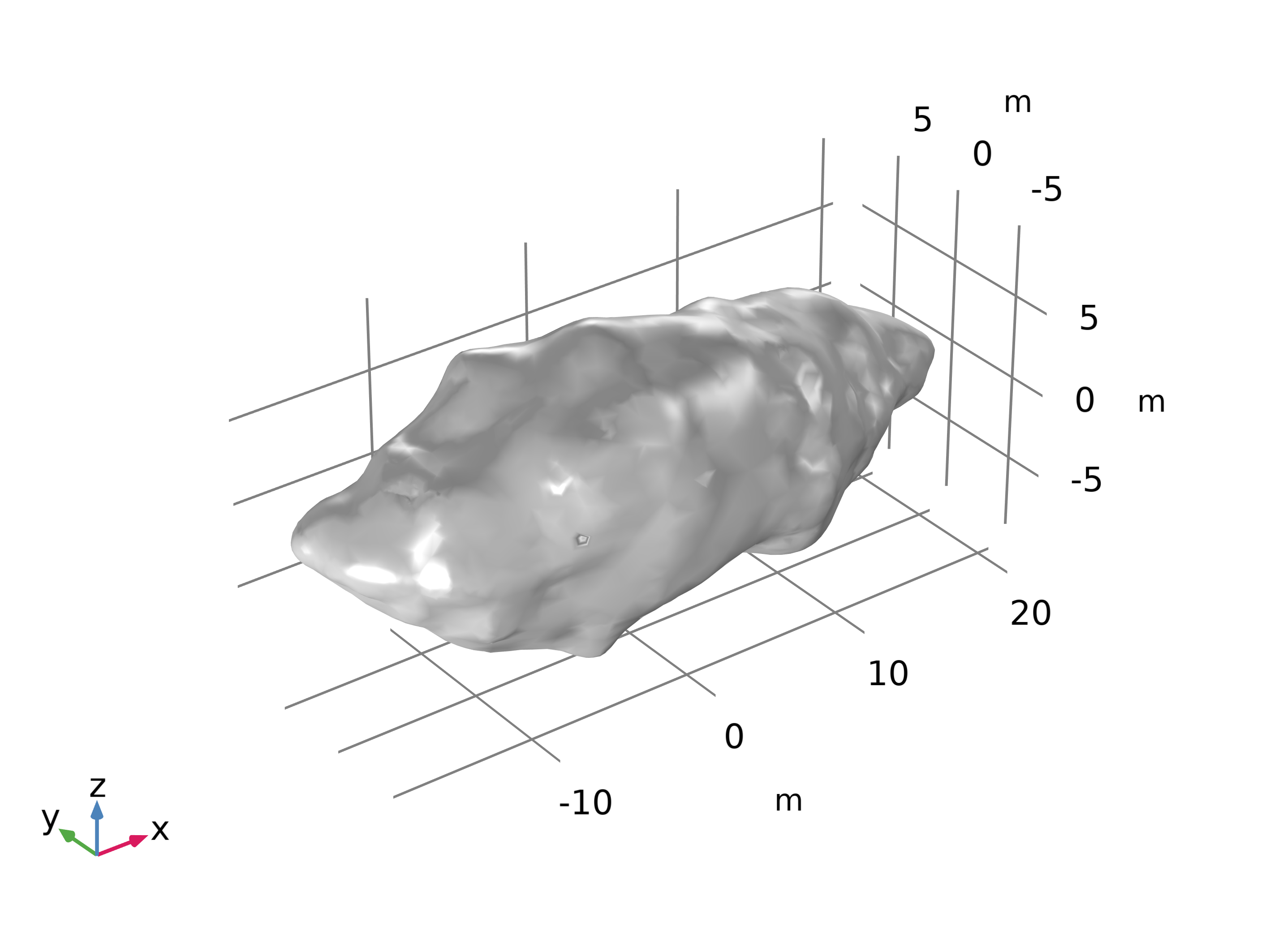}
		\put(2,70){(b)}
	\end{overpic}
	\caption{The shapes of asteroids (a) 1994 KW4 Beta and (b) Geographos. }
	\label{fig:kw4geog}
\end{figure}

Asteroids 1994 KW4 Beta and Geographos (Fig.~\ref{fig:kw4geog}) are good examples to show the intuitive connection between the shape index and their shape. Approximately regard them as ellipsoids, the 1994 KW4 Beta with dimensions ($0.571\times 0.463\times 0.349$)\,km has $e_1 = 0.68, e_2 = 0.81$, and Geographos of ($5.0\times 2.0\times 2.1$)\,km has $e_1 = 0.66, e_2 = 0.4$ \citep{2006Sci...314.1276O,1999Icar..140..369H}. They have similar $e_1$ but quite different $e_2$ values. The shape indices of these two ellipsoids are $0.811$ and $0.954$. The values calculated from the real shape models are $S_1=0.82$ ($S_2=0.81$) for 1994KW4 Beta and $S_1=1.10$ ($S_2=1.06$) for Geographos, respectively, agreeing with the ellipsoid approximation.

In addition, top-shaped asteroids are quite common, with Ryugu and Bennu being the ready examples. A top-shaped asteroid can be reduced to a combination of two cones with a common base and oppositely directed tops. Denote the ratio between the equatorial radius $R$ and height $h$ of the northern and southern cones by $r$, that is, $r = R/h$, we can write the shape index $S_1$ as:
\begin{equation}
	S_1\left(r\right)=\frac{3}{4}\left(\frac{2}{r}\right)^{1/3}\left(\frac{1}{1+r^2}\right)^{1/2}.
	\label{eq:cones}
\end{equation} 
Clearly, $S_1$ decreases monotonically as $r$ increases. We find that $S_1(r=0.56)=1$, implying that this top-shaped asteroid has equivalent Yarkovsky effect as a sphere asteroid when the cone angle is $\sim$30$^\circ$. Otherwise, when the cone angle is smaller than 30$^\circ$ ($r<0.56$), the object will have a relatively stronger Yarkovsky effect and the shape index $S_1>1$. For Ryugu and Bennu, their equatorial to polar axis ratios of 1.147 and 1.104 \citep{2019Sci...364..268W,2019Natur.568...55L} give $S_1=0.593, 0.614$, respectively. These values are much smaller than their actual shape indices $S_1=0.955, 0.960$ calculated from their real shapes, implying that the jointed-cones is not a good model for these two asteroids. In fact, some ``plains'' can be found in the equatorial region of both asteroids, and they might contribute considerably to the real shape index.     

So far, the cylinders, ellipsoids and spinning top shape discussed above are all fully convex, which means $S_1 = S_2$ for them. In the real world, the boulders and craters on the surface of asteroids may increase the specific surface area, alter the orientation of surface elements, and produce shadows. Although the influences of these irregularities on shape index are highly complex, a general estimation can still be made by approximating the real shape as a deformed version of its inertia ellipsoid. For the 34 shape models, we calculate the shape index both for real shapes ($S_2$ ) and for their inertia ellipsoids ($S'_2$), and find that for almost every model the $S_2$ of the real shape is larger than that of its inertia ellipsoid. Especially for those of very irregular shapes, their $S_2$ are significantly larger than the corresponding $S'_2$. Again, taking the four asteroids in Fig.~\ref{fig:S2VarTheta} as examples, we listed in Table~\ref{tab:realellip} the two shape indices. We note the difference between $S_2$ and $S'_2$ for Kleopatra is over 20\%.  
 
  \begin{table}
 	\centering
 	\caption{The shape indices of four shape models ($S_2$ ) and their inertial ellipsoids ($S'_2$ ).}
 	\begin{tabular}{c | c c c c}
 		\hline
 		\hline
 		~  & YORP & Kleopatra & Ida & Golevka\\ 
 		\hline
 		 $S_2$  & 0.993 & 1.167 & 0.889 & 1.073 \\
 		 $S'_2$ & 0.875 & 0.943 & 0.753 & 0.937 \\
 		\hline
 	\end{tabular} \label{tab:realellip}
 \end{table}
 
Roughly speaking, data in Table~\ref{tab:realellip} implies that the irregularities in shape are likely to increase the shape index, and a stronger Yarkovsky effect might be expected if the asteroid is very irregularly-shaped.  

\section{Conclusion}
\label{sec:concls}
The Yarkovsky effect plays an important role in the orbital evolution of asteroids. In addition to the thermal, physical, and dynamical parameters, the Yarkovsky effect of an asteroid strongly depends on its shape, which brings great complexity to both the analytical and numerical calculations of the effect. To obtain a quick estimation of the diurnal Yarkovsky effect, \citet{xu2022diurnal} proposed the idea of ``effective area'', which can be easily calculated if the shape of an asteroid is known. In this paper, we improve the idea and refine an index that can quantitatively measure the influence of asteroids' shape on the Yarkovsky effect. 

We adopt 34 real asteroids with shape data available as the sample of shape models, rescale the sizes of these models so that they all have the same volume as a sphere with a radius of 10 meters, assign them with specific thermal parameters, and put them all in a circular orbit with semi-major axis of 1\,AU. We established the numerical model in the COMSOL software package to calculate the temperature on their surface, from which the Yarkovsky force and then the semi-major axis drift rate ($\dif a/\dif t$) can be derived. The migration rate $\dif a/\dif t$ was used to measure the strength of the Yarkovsky effect. 

After revisiting the definition of effective area in \citet{xu2022diurnal}, which is roughly a double projection of the surface area onto a plane parallel to the spin axis, we normalized it to a shape index $S_1$ (Eq.\eqref{eq:s1}). When defining $S_1$, we have assumed that all energy absorbed by a surface element will be released by thermal radiation after a delay angle $\theta$ from the same surface element. For a given asteroid, the Yarkovsky effect is determined by the dimensionless thermal parameter $\Theta$. Although the delay $\theta$ can be calculated if $\Theta$ is known, we proved that the shape index $S_1$ does not depend on the delay $\theta$ specifically. We also showed by numerical simulations that the assumption of all surface elements having the same delay angle is valid (Fig.~\ref{fig:angle}). Therefore, this shape index $S_1$ is determined solely by the shape of an asteroid, and a linear relationship between $\dif a/\dif t$ and $S_1$ as proposed in  \citet{xu2022diurnal} was reaffirmed using a larger sample of asteroids (Fig.~\ref{fig:S1S2}). 

We found that the outliers in the fitted linear relationship are those that might be most influenced by the shadowing effect. We therefore improved the shape index by removing the contributions from  surface elements when they are under the projected shadows. The linear relationship between $\dif a/\dif t$ and the improved shape index $S_2$ (Eq.\eqref{eq:s2}) is much better than the former one with respect to $S_1$, and the outliers are tamed (Fig.~\ref{fig:S1S2}). As for the linear fit, the coefficient of determination ($R^2$) increases from 0.984 for $S_1$ to 0.995 for $S_2$. 

When defining the shape index, a 1D heat transfer model is assumed and we totally ignore the lateral heat transfer. This assumption might be violated if the asteroid has a large thermal conductivity. We adopted three conductivity values in a wide range over four orders of magnitude ($K_1=0.0015, K_2=0.15, K_3=15$\,W\,m$^{-1}$\,K$^{-1}$), and tested the performance of shape index $S_2$. We found that the strength of Yarkovsky effect changes remarkably as the thermal parameter $\Theta$ varies for different $K$, but the linear relationship between $\dif a/\dif t$ and $S_2$ is robust (Fig.~\ref{fig:S2VarTheta}).

We extended the calculations and obtained the fitting parameters (the slope $k$ and intercept $b$) of the linear relationship for other $\Theta$ values. As the $\Theta$ increases, the slope $k$ increases first and then decreases, reaching its maximum when $\Theta\sim1.74$. The linear relationship is very well kept for all $\Theta$ values. With these calculations, if the thermal parameter $\Theta$ is known, the Yarkovsky effect of an asteroid of any shape can be directly estimated by its shape index $S_2$ using the linear relationship with the parameters given in Fig.~\ref{fig:knb}.

The rate of heat transfer inside an asteroid depends on both the thermal conductivity and temperature gradient. An increase in $K$ (thus in $\Theta$) results in quicker heat diffusion inside the body and more evenly distributed temperature over the surface (Fig.~\ref{fig:tempdis}) that often leads to a weaker Yarkovsky effect. Furthermore, a higher $\Theta$ reduces the temperature gradient along the east-west direction much more than it does along the north-south direction (Figs.~\ref{fig:tempdis}, \ref{fig:equatemp}). We argue that the lateral heat conduction inside an asteroid is not significant, even when it has a high thermal conductivity. This was also confirmed by the negligible difference of Yarkovsky effect between a 1D model (without the lateral transfer) and a 3D model (with the lateral transfer) for the same spherical body. Therefore, the estimation of Yarkovsky effect by its linear dependence on the shape index $S_2$ is valid for all bodies satisfying $R\gg l_{\rm d}$. 

The radiation scattering and the self-heating effects may affect the Yarkovsky effect and their influences may be enhanced when the thermal conductivity increases. However, through numerical simulations, we found that the contributions from these effects to the Yarkovsky effect are ignorable. Even for very irregularly-shaped asteroids with very large $K_3=15$\,W\,m$^{-1}$\,K$^{-1}$ (thus large $\Theta=73.6$) value, their contributions makes only a relative difference in the Yarkovsky force of $\sim$10$^{-3}$ (Table~\ref{tab:self-heat}). After incorporating these effects into the definition of shape index (now $S_3$), we found no improvements to the linear relationship between $\dif a/\dif t$ and $S_3$ (Table~\ref{tab:R2}). In fact, without knowing the details of surface temperature, the influence of self-heating cannot be included precisely in the estimation of Yarkovsky effect. Therefore, we conclude that it is not necessary to take into account the scattering and self-heating effects in our definition of shape index.  

To demonstrate the dependence of shape index on specific shapes, we computed the $S_1$ and $S_2$ for several sets of simple shape models, including cylinders, ellipsoids (Fig.~\ref{fig:ellipS1}) and top-shaped asteroids (Eq.\,\eqref{eq:cones}), and found that prolate shapes, such as ellipsoids with oblateness $e_1>1$ and spinning tops with equatorial to polar axis ratio $r<0.56$ would have large shape index $S_1>1$, while the oblate shapes have smaller $S_1$. In addition, for irregularly shaped asteroids, we calculated the $S_2$ both for their real shapes and for their inertia ellipsoids (Table~\ref{tab:realellip}), and found that irregularities in shape generally increase the shape index $S_2$.

As we have shown, the shape index ($S_2$) is a good indicator of the strength of diurnal Yarkovsky effect. It is solely determined by the shape of an asteroid, and can be easily computed, bringing great convenience to the estimations of Yarkovsky effect for a large number of asteroids. Knowing the shape of an asteroid by observations, the Yarkovsky effect can be easily obtained by comparing its shape index with the one of an ``ideal standard asteroid'' (e.g. a spherical asteroid) with the same thermal parameters. Meanwhile, the shape of an asteroid is generally inverted from its light curves. With the shape index as a bridge, we may be able to find a direct link between the Yarkovsky effect of an asteroid and its light curves, on which we are working through machine learning now.

\begin{acknowledgements}
This work has been supported by the National Natural Science Foundation of China (NSFC, Grants No.12373081 \& No.12150009) and the China Manned Space Program with grant No.CMS-CSST-2025-A16. We also acknowledge the support from National Key R\&D Program of China (2019YFA0706601). The temperature distribution on the asteroid's surface is computed using COMSOL software. %(license No.9409940).
\end{acknowledgements}
	
\bibliographystyle{aa.bst}
\bibliography{reference.bib}

\end{document}